# Compaction behaviour of various flax fabric structures during composite manufacturing: mechanical characterisation and microstructural analysis

R. Rayyaan[1, 2*], Z. Yousaf[1, 3*], W. R. Kennon[1], P. Potluri[1]

[1]Department of Materials, University of Manchester, Oxford Road, Manchester, M13 9PL, UK

[2]Department of Education, King's College London, UK

[3]Department of Mathematics, University of Manchester, Oxford Road, Manchester, M13 9PL, UK

**Abstract**

Flax fibre reinforced composites are becoming popular in the automotive and civil industries due to their environmentally friendly nature in terms of production and recycling and for their good specific strength. Textile structures undergo transverse compaction during composite manufacturing which changes the fabric thickness and ultimately fibre volume fraction of composites. Here, different flax fibre structures were investigated for compaction behaviour during composite forming to study the thickness changes of these fabrics under pressures between 1 and 10 bars. A range of composite manufacturing processes including vacuum infusion, autoclave curing and resin transfer moulding were utilised. These fabrics were studied in single and multi-layer states, in dry and wet states, under different loading cycles and in different orientations of the fabric plies ($0^0/0^0$ and $0^0/90^0$). Nesting of the layers has been calculated for single plies and for multi-layer stacks of dry fabrics. It was observed that under transverse compression, the structure of the preform plays a vital role in determining the thickness of the preform. The compressibility pattern of these various structures was also dissimilar for single layer and multi-layer stack which was attributed to the different nesting behaviours of these flax fibre structures.

*Corresponding authors R. Rayyaan (rishad.rayyaan@kcl.ac.uk),
 Z. Yousaf (zeshan.yousaf@manchester.ac.uk)

# 1. Introduction

Flax fibre is looming large for use in structural composite components, mostly because of its superior mechanical properties within the domain of natural fibres [1] and environmental friendly nature. In terms of initial longitudinal modulus of composites, flax-reinforced composites can produce performances as good as glass-reinforced composites [2]. However, the translation efficiency of fibre-properties to composites' properties still needs to be improved in order to best exploit the load-bearing potential of this fibre. When flax fibres are turned into preforms and are stacked into layers prior to composites manufacturing, the compaction behaviour of the preforms plays an important role in determining the composites' fibre volume fraction.

A number of different techniques for the manufacturing of advanced polymer composites such as vacuum Infusion (VI), autoclave curing, resin transfer moulding (RTM) etc. are popular [3, 4]. In these processes, the textile fabrics are compacted to an extent which results in a change in the thickness and fibre volume fraction of the preforms. Compaction of the fabrics also changes the voids for the resin flow and the mechanical properties of the final product, so it is taken as an important parameter of the manufacturing process [5]. The compaction of a single layer fabric is different from that of multi-layer fabrics. Compaction of multi-layer preforms involves the nesting of the layers due to shifting of the individual layers, resulting from pressure upon the preform stack. The nesting of the layers in a multi-layer stack changes the thickness of the stack; hence the final fibre volume fraction of the resulting composite changes [6-9]. Therefore, study of the nesting phenomenon is important to understand the compaction of textile preforms.

In published literature, the compaction behaviour of glass and carbon fabrics is widely described [10-15] but relatively little research can be found on the compaction of flax fabrics [16, 17]. It is thus necessary to investigate the thickness and microstructural changes of flax preforms under loading in the dry and wet states during compaction whilst creating bio-composites. In this research, compaction behaviour of different flax fibre structures has been examined during composite forming to study the thickness changes of these fabrics under variable pressure to simulate different composite manufacturing processes including vacuum infusion, autoclave curing and resin transfer moulding. The fabrics have been observed as single layers and as multi-layers, under dry and wet states, under different loading cycles, and with the fabric plies oriented in different directions ($0^0/0^0$ and $0^0/90^0$). Nesting of the layers has been calculated for single plies and multi-layer stacks of dry fabrics. And three different methods of calculating the fibre volume fraction of composites have been described.

## 2. Material and mechanical testing

### 2.1 Material

Four types of fabric have been used in this research namely Hopsack (plain woven hopsack fabric), UD (unidirectional warp knitted fabric made from twistless wrap-spun yarn), T170 (nonwoven tape of 170 mm width), and GVT (nonwoven tape covered with a glass fibre veil). Details of the materials used in this study are given in Table 1.

**Table 1  Dry fabric specifications**

|  | Hopsack | UD | T170 | GVT |
|---|---|---|---|---|
| **Yarn Linear density, in tex** | 250 | 250 | n/a | n/a |
| **Flax content, %** | 88 | 84.53 | 90 | 89.5 |
| **Glass content, %** | n/a | n/a | n/a | 10.5 |
| **Polyester content, %** | 12 | 15.47 | n/a | n/a |
| **PLA content, %** | n/a | n/a | 10 | n/a |
| **Fabric construction** | 4×4 hopsack | Warp knitted | Nonwoven tape | Nonwoven tape with surface veil |
| **Areal density, in g/cm2** | 0.0519 | 0.0264 | 0.0158 | 0.049 |
| **Fabric density, g/cc** | 1.4845 | 1.4802 | 1.467 | 1.564 |
| **Ends/inch** | 24 | 24 | n/a | n/a |
| **Picks/inch** | 30 | 6 | n/a | n/a |

The top views and cross sectional images of the fabrics used in this study obtained by optical microscopy are presented in the photomicrographs of Figure 1&2 Figure respectively. In addition to flax fibre, the cross section of a flax elementary fibre is also presented in Figure 3. It can be seen from Figure 3 that the flax fibre has a hollow shape inside the fibre. The permeability deriving from the hollow shape of the flax fibre can play a role during the infusion of flax fibre composites.

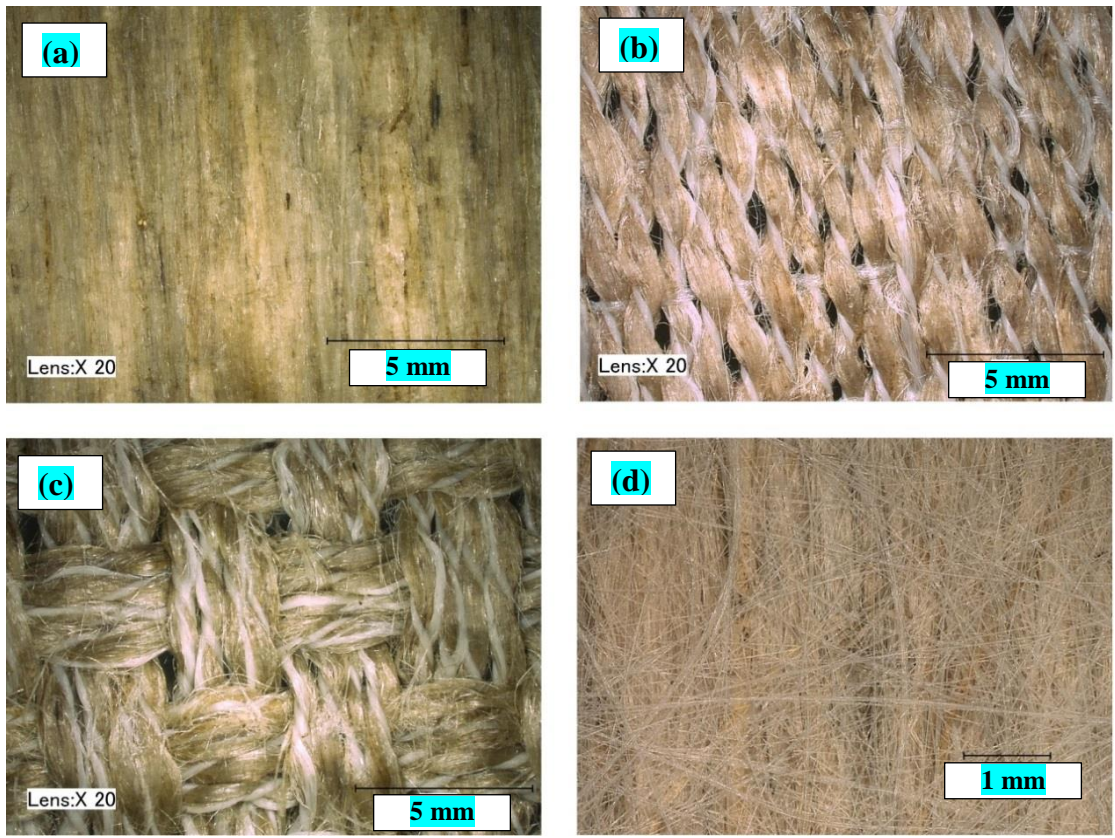

**Figure 1    Optical images of flax fabrics (top view) (a) Nonwoven tape; (b) Warp knitted UD fabric; (c) Hopsack fabric; (d) Nonwoven tape with surface veil.**

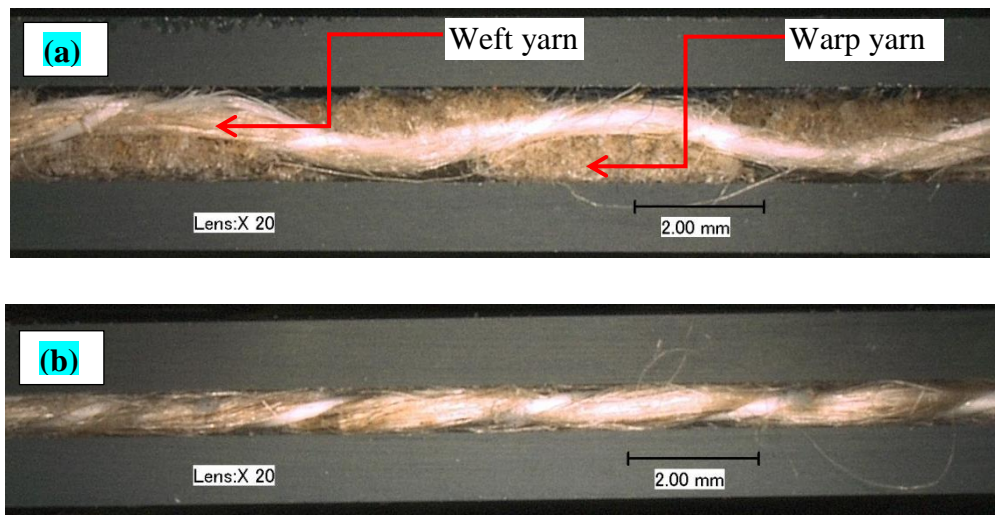

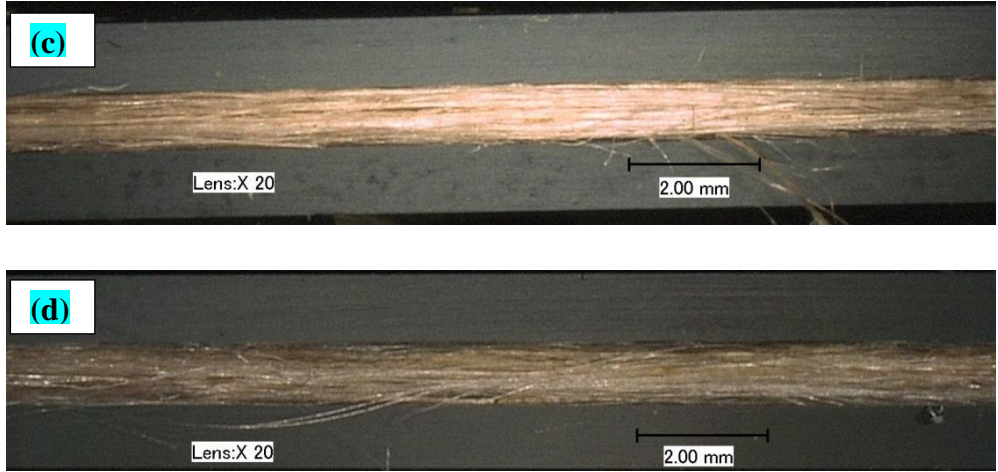

**Figure 2** Cross sectional images of flax fabrics captured by optical microscope: (a) Hopsack fabric; (b) UD; (c) T170; (d) GVT.

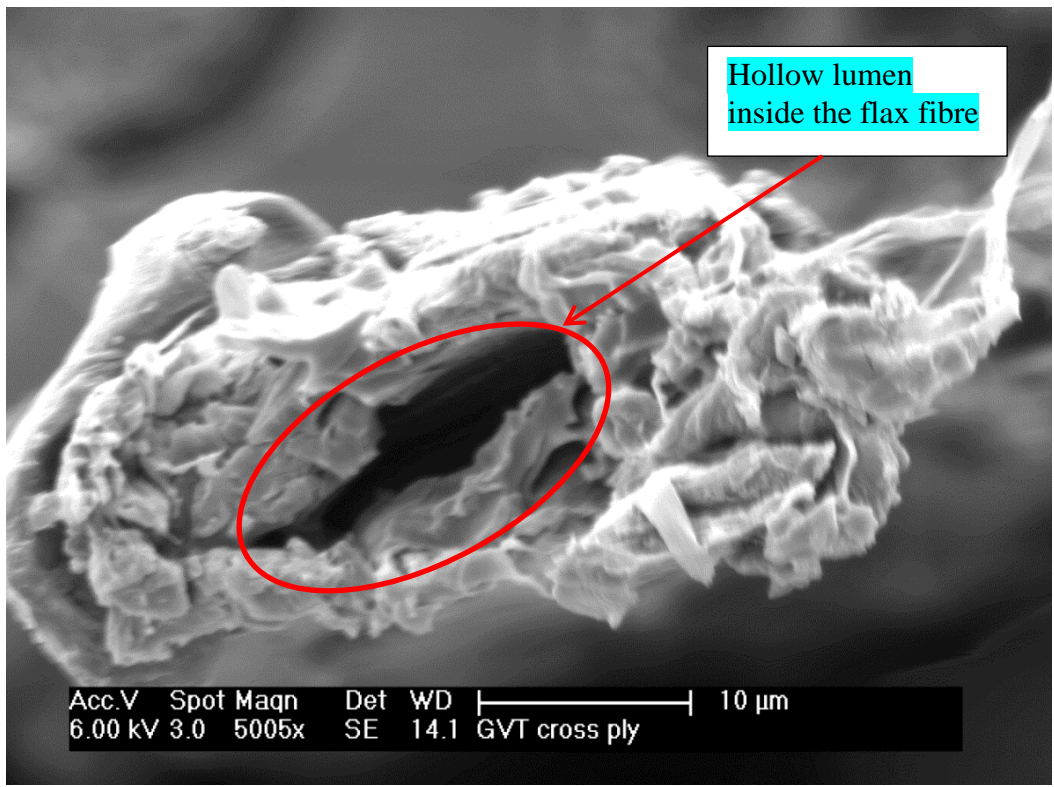

**Figure 3** Elementary flax fibre; cross-sectional SEM image showing lumen.

**2.2 Compaction testing**

Compaction testing was done using an Instron 5569 universal testing machine as depicted in Figure 4. On the Instron 5569, two circular metal plates were used to define the compaction of the woven fabric samples. The area of the top and bottom plates was 50 cm$^2$. The samples were cut into 100 mm × 100 mm pieces. Testing of the samples was executed using a dynamic method. Under dynamic control, the machine was moved continuously at a constant speed of 1 mm/min and the pressure/thickness curve was recorded. Before the start of the testing, correction was performed to remove any error due to machine compliance.

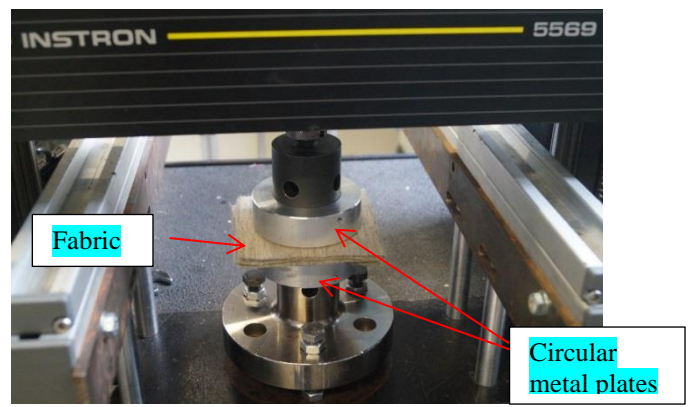

**Figure 4    Instron testing set up for fabric compression.**

**3.  Results and Discussion**

**3.1 Macroscopic deformation of single and multi-layer structures**

The graph in Figure 5 shows the thickness of various single layer flax fibre structures when subjected to increasing pressures up to a maximum of 10 bars, applied by the Instron testing machine. It is clearly evident that the thickness of all the structures decreased with increase in pressure and the behaviour of the curves was the same as predicted by the pressure-thickness curve in published research [18].

The greatest reduction in thickness in response to increase in pressure was observed in the range up to 100 kPa, which is representative of vacuum infusion. While there was less reduction in thickness of all the structures beyond that pressure, the initial thickness of the GVT fabric was highest and T170 was the lowest. Initial thickness can be mainly attributed to the areal density. The sequence of the areal density of the specimens, from low to high, is: T170<UD<Hopsack<GVT, as shown in Table . Recording of the thickness for Figure 5 has been started from a load of 0.98 N, which is equivalent to a 100 g load on the fabric. At 0.98N (0.18kPa) load, the sequence of thickness from low to high is: T170<UD<Hopsack<GVT, which still follows the order of areal density. Table shows the thickness reduction percentage of each specimen. During an increase of load from 5 kPa to 100 kPa, the greatest thickness reduction was observed for GVT, and this was 40.1%. The other three structures showed approximately equal reductions in thickness; close to 30%.

The reason for the GVT to have shown greatest thickness reduction can be attributed to its glass veils. Under the pressure exerted by the steel plates during compaction, the fine glass fibres in the veils are crushed which results in a high percentage reduction of the fabric thickness.

**Table 2   Thickness reduction for single layer flax fabrics**

|   | Thickness in mm (at 1 kPa) | Thickness in mm (at 100 kPa) | Thickness Reduction (%) |
|---|---|---|---|
| **T170** | 0.46845 | 0.32445 | 30.74 |
| **UD** | 0.7119 | 0.49963 | 29.82 |
| **Hopsack** | 1.34527 | 0.94636 | 29.65 |
| **GVT** | 1.6274 | 0.97476 | 40.10 |

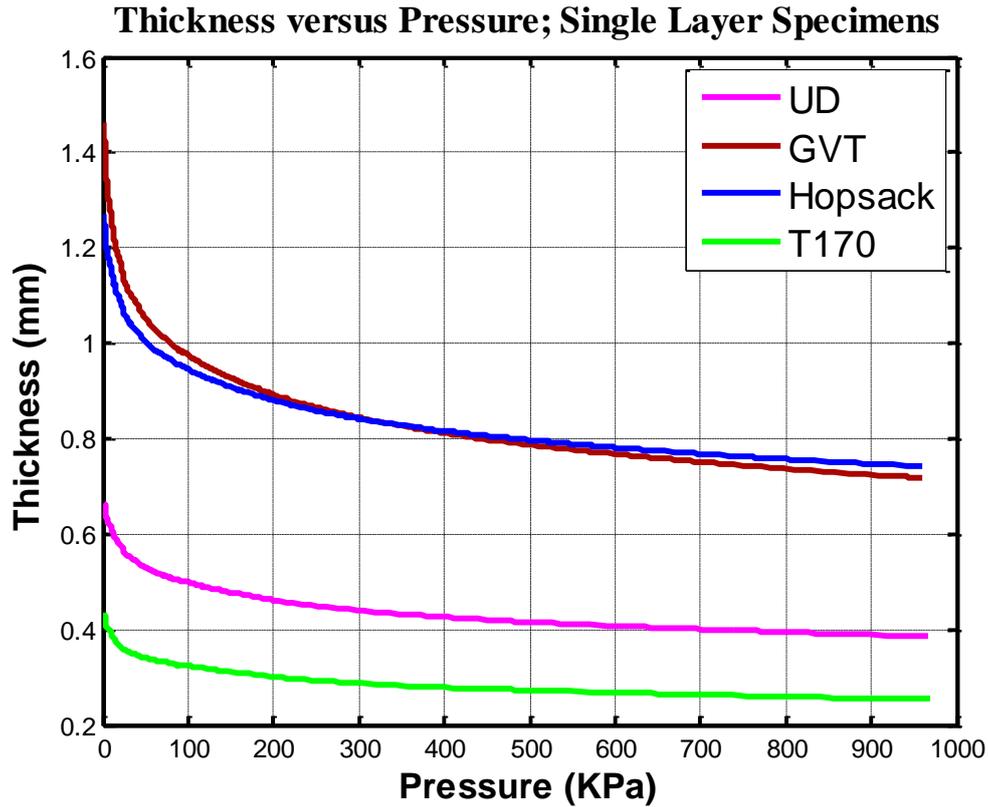

**Figure 5** Pressure vs thickness responses of single layer flax fibre structures.

Figure 6 shows the pressure-thickness curves of various multi-layer flax fibre fabric specimens. All the specimens were laid-up in a cross-ply manner (0/90°) with each structure having 4 layers. The pressure-thickness curves of the stack of multi-layered specimens also followed the same behaviour as depicted by a typical pressure-thickness curve in previously published scientific literature [18].

**Table 3** Thickness reduction for multi-layer flax fabrics

|  | Thickness in mm (at 1 kPa) | Thickness in mm (at 100 kPa) | Thickness Reduction (%) |
| --- | --- | --- | --- |
| **T170** | 2.12 | 1.37 | 35.45 |
| **UD** | 2.86 | 1.95 | 31.63 |
| **Hopsack** | 5.27 | 3.54 | 32.82 |
| **GVT** | 6.86 | 4.18 | 39.02 |

Table 3 shows the thickness reduction percentages for the multi-layer stacks. It can be seen that the greatest thickness reduction for multi-layer stacks has once again been demonstrated by the GVT structure, echoing its behaviour as a single layer.

While the GVT specimens underwent a 39% thickness reduction, the Hopsack and the UD specimens suffered a 31% and 32% thickness reduction respectively. However, the T170 experienced a thickness reduction of 35%. As previously, the greatest percentage thickness reduction was shown by GVT and may be attributed to the distortion of the glass veils and the lower nesting tendency of the GVT specimens prior to loading. In addition, the individualisation of the technical fibres into elementary fibres may also contribute to this higher thickness reduction percentage; a detailed discussion is presented in section 3.3.

The thickness reduction percentage of the T170 is greater than that of the UD or Hopsack structures but lower than that of the GVT. The higher value for T170 (35%) compared with those of Hopsack and UD (31% and 32%, respectively) can be attributed to the individualisation of the technical fibres and the reduced nesting of the T170. Nesting factors are discussed in detail in sections 3.2 and 3.3.

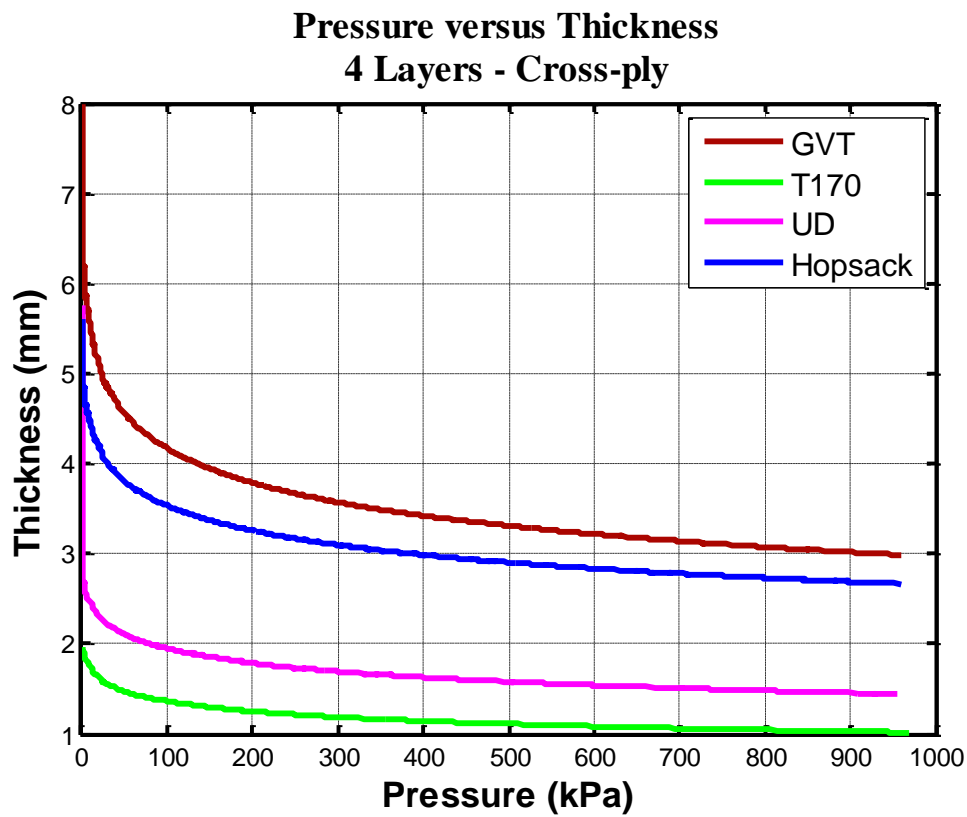

**Figure 6**  Pressure vs thickness curves of various flax fibre structures in multi-layered stacks.

From the thickness-pressure curves of both the single layer and the multi-layer stacks, it can be extrapolated that nonwoven structures demonstrate a higher percentage of thickness reduction in comparison with woven structures. Therefore, in order to obtain a composite panel of a specific thickness, a composite structure will require more nonwoven flax fabric layers than when using woven flax fabric layers.

### 3.2 Nesting factor

The nesting of layers in a multi-layer stack changes the thickness and hence the fibre volume fraction of the textile preform. The nesting factor can be calculated by employing equation 1.

$$NF = T_{avg} / T_s \qquad (1)$$

Where, NF is the nesting factor, $T_{avg}$ is the average layer thickness in a multi-layer stack and $T_s$ is the thickness of the single layer. In this work, the single and multi-layer fabric thicknesses were calculated from the mechanical test results of these fabrics and the nesting factors were calculated by applying equation 1. Table 4 shows the nesting factors for all the specimens. Figure 7 shows graphs of the thicknesses of the single layers and of the averaged layers. An 'averaged layer' thickness has been calculated from the multi-layer stack, dividing the multi-layer stack thickness by the number of plies. Here, all the multi-layered specimens have been laid-up in a 0/90° orientation, using 4 plies for each type.

**Table 4   Nesting factor (NF) for all the specimens**

| Thickness (mm) Pressure (KPa) | | Hopsack | | UD | | GVT | | T170 | |
|---|---|---|---|---|---|---|---|---|---|
| | | Single layer | Multi (4layers) | Single layer | Multi (4layers) | Single layer | Multi (4layers) | Single layer | Multi (4layers) |
| 100 | Thickness | 0.9464 | 3.5396 | 0.4984 | 1.9535 | 0.9748 | 4.1829 | 0.3245 | 1.3678 |
| | NF | 0.9351 | | 0.9799 | | 1.0728 | | 1.0540 | |
| 200 | Thickness | 0.8800 | 3.2579 | 0.4598 | 1.7866 | 0.8917 | 3.7880 | 0.2998 | 1.2515 |
| | NF | 0.9255 | | 0.9714 | | 1.0620 | | 1.0437 | |
| 400 | Thickness | 0.8164 | 2.9880 | 0.4252 | 1.6283 | 0.8118 | 3.4264 | 0.2783 | 1.1434 |
| | NF | 0.9150 | | 0.9575 | | 1.0552 | | 1.0271 | |
| 800 | Thickness | 0.7572 | 2.7305 | 0.3940 | 1.4805 | 0.7370 | 3.0754 | 0.2595 | 1.0442 |
| | NF | 0.9016 | | 0.9395 | | 1.0433 | | 1.0058 | |
| 1000 | Thickness | 0.7375 | 2.6522 | 0.3837 | 1.4342 | 0.7124 | 2.9671 | 0.2529 | 1.0141 |
| | NF | 0.8990 | | 0.9345 | | 1.0412 | | 1.0024 | |
| 1200 | Thickness | 0.7232 | 2.5889 | 0.3746 | 1.3967 | 0.6934 | 2.8802 | 0.2476 | 0.9875 |
| | NF | 0.8950 | | 0.9320 | | 1.0385 | | 0.9969 | |
| 1400 | Thickness | 0.7099 | 2.5362 | 0.3670 | 1.3623 | 0.6766 | 2.8078 | 0.2433 | 0.9681 |
| | NF | 0.8931 | | 0.9279 | | 1.0375 | | 0.9948 | |

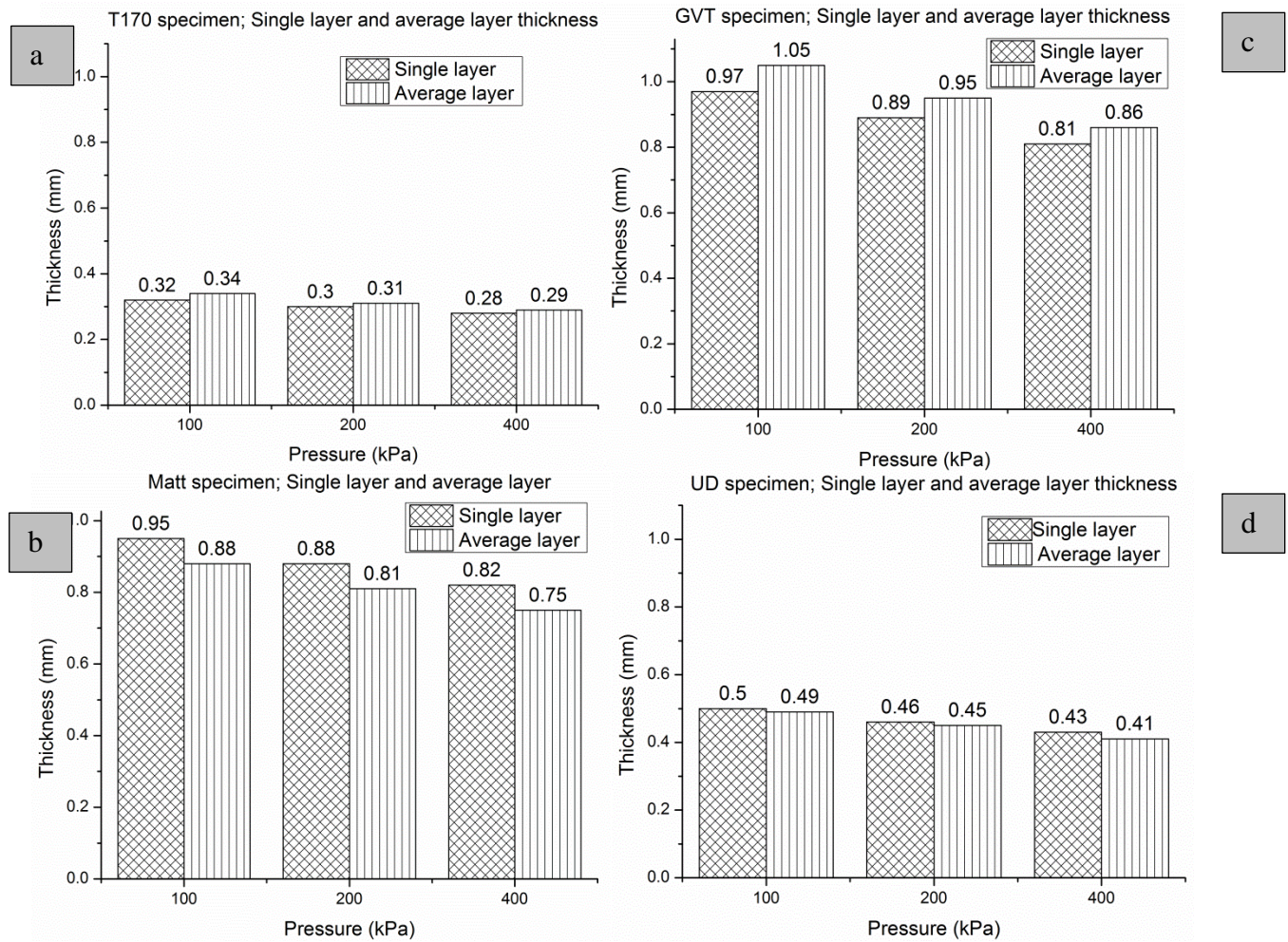

**Figure 7　Single and averaged layer thicknesses of different flax fabric architectures under various pressures.**

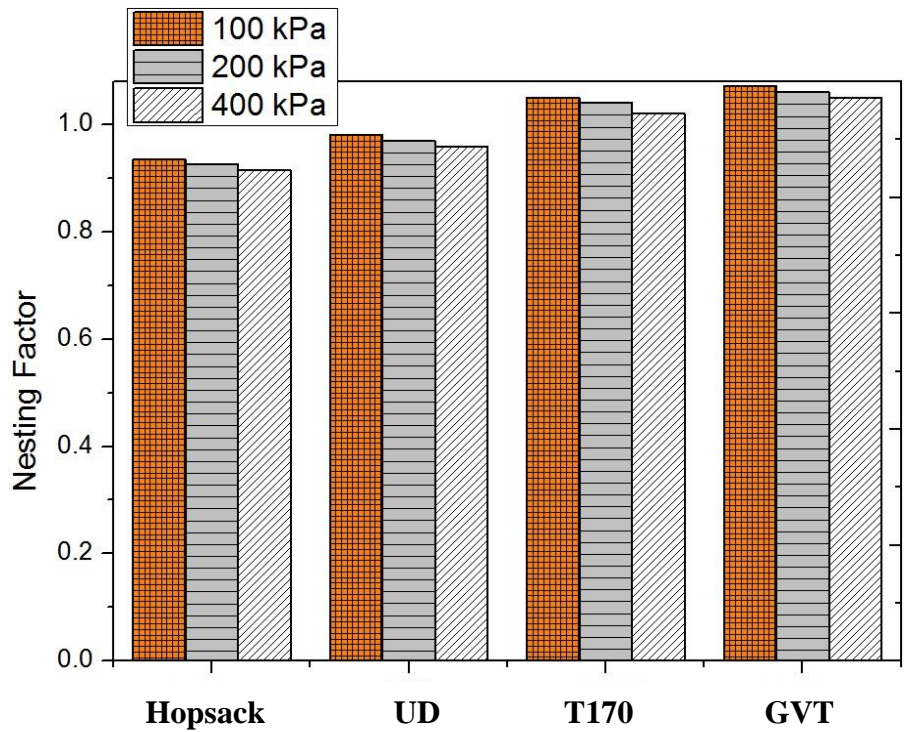

**Figure 8**  Nesting factors of different flax fabric architectures.

A lower nesting factor indicates better nesting and vice versa. The average layer thicknesses of multi-layer stacks and single layer thicknesses of flax fabrics calculated from mechanical test results at different pressures are presented in Figure 7, whereas Figure 8 represents the nesting factors calculated at pressures of 100, 200 and 400kPa from the single layer thickness results combined with the average stack thickness results obtained by applying equation 1. It can be seen that the nesting factors decreased with increase in pressure for both woven and nonwoven structures, which means better nesting efficiency of the multi-layer stack.

**3.3 Analysis of the nesting factor**

Nesting is a normal phenomenon that occurs during the compaction of multiple layers of fabric, especially for woven fabrics. The yarn in a 2D or 3D woven fabric follows an undulating path which leaves some gaps between successive parallel layers and between successive layers normal to each other (warp and weft fabrics) when in a stack. During composite manufacturing, the fabric plies come into contact under pressure. For the vacuum assisted resin infusion process, the pressure happens to be around 1 bar (100 KPa). For autoclave processing, the applied pressure could reach 9 bars or more [19]. Under such pressures, the warp (or weft) yarn of one ply tends to set into the gaps between the two adjacent warp yarns of the ply above or beneath. A simple demonstration with two plies of dry hopsack fabric sandwiched between two microscope slides with a nominal pressure applied by a pair of binder clips is shown in Figure 9.

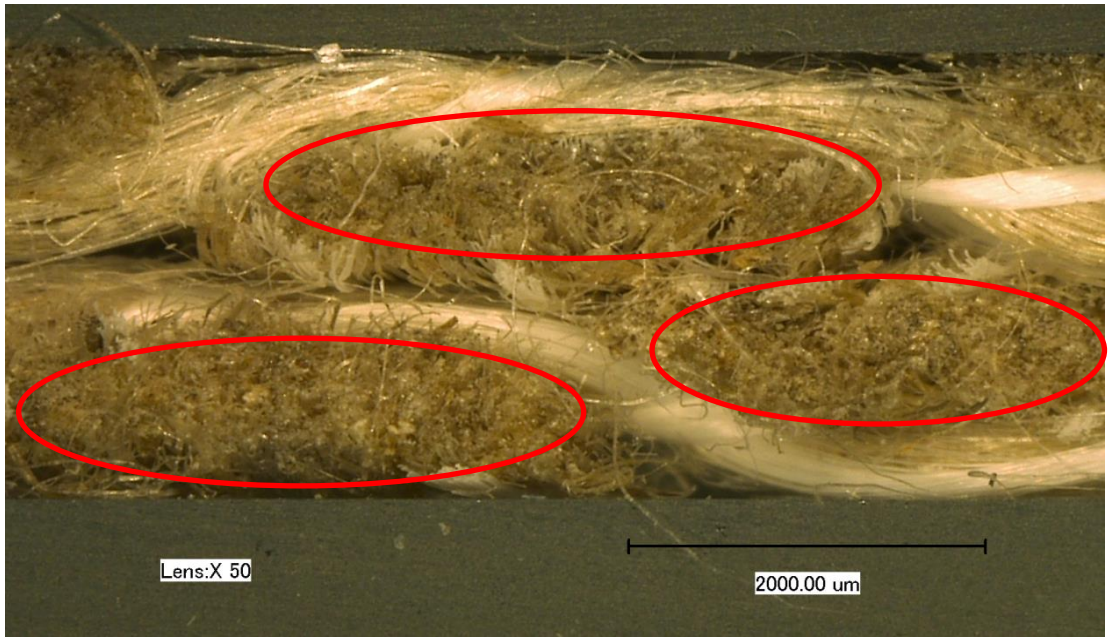

**Figure 9    Two layers of Hopsack fabric; showing nesting between adjacent layers**

In the photomicrograph of Figure 9, three warp yarns are bordered with red lines; each ellipse indicates a warp yarn. It can be seen that the top yarn (in the middle) has settled down into the gap between the two yarns from the under layer.

Normally, the thickness of a single layer is found to be greater than the average thickness of that single layer when stacked as a component of a multi-layered composition. This situation occurs because within a multi-layer structure, one layer enters into the planes of the adjacent layers to some degree. The overall thickness of a multi-layer stack can be estimated by simply multiplying the thickness of a single layer by the number of laminates used to form the multi-layer. But in reality, the overall thickness of the multi-layer stack tends to be slightly lower than the thickness value estimated by multiplication (number of layers in the multi-layer $\times$ thickness of a single layer). This reduction in the multi-layer thickness value occurs due to nesting, which generates a nesting factor value less than 1.

Nevertheless, exceptions to this case can also sometimes occur. Pearce and Summerscales (1995) have reported that the nesting factor can occasionally be greater than 1. They have explained the reason is that a single layer undergoes compression between two polished steel platens during compaction testing. The friction between the steel plates and the fabric tends to be lower than the friction between two fabric layers in a multi-layer stack. The steel plates flatten a single layer to a certain extent. But when that layer resides as a middle layer in a multi-stack, the layers above and beneath cannot flatten the middle layer as much as the steel plates can, because of the higher frictional force between the adjacent fabrics. So, the cohesive force between the two fibre surfaces plays a role in holding the fibres in contact. Eventually, not all the peaks and troughs of one layer can be adjusted into the undulations of the adjacent layers. As a result, each layer contributes a greater thickness in an arrangement of a multi-layer stack compared with that of the sole thickness of a single layer [20].

The probability of having a nesting factor higher than 1 (one), explained by Pearce and Summerscales (1995), describes the woven fabric compaction behaviour. Interestingly, in this research, the nesting factor has been calculated as higher than 1 for the nonwoven tapes (T170), and also for the nonwoven tapes with glass veils (GVT).

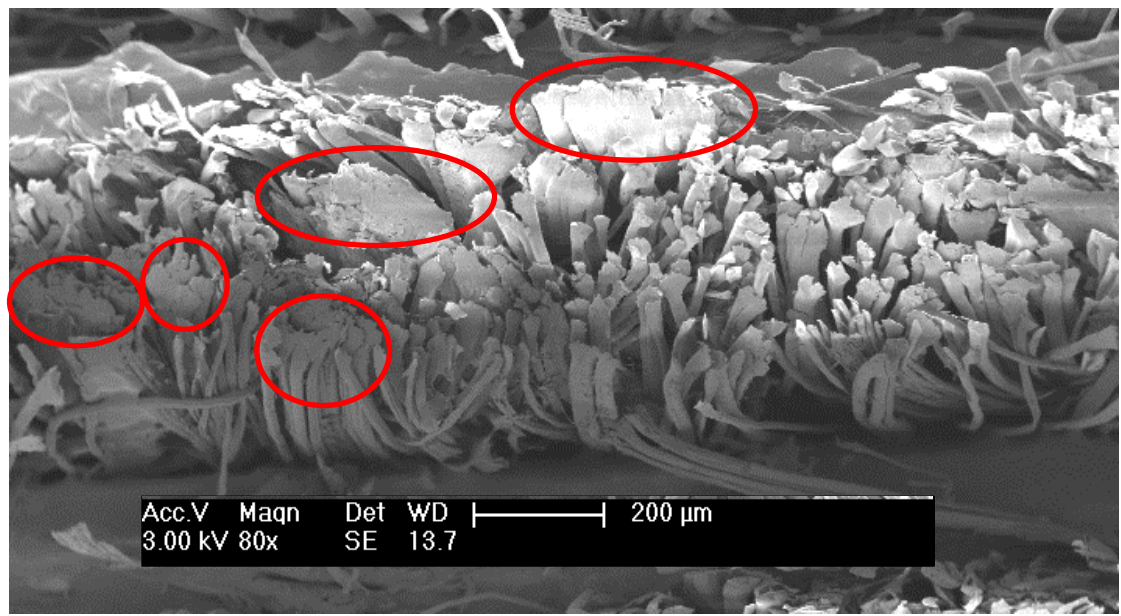

**Figure 10   Highlighted undamaged fabric-1; clusters of elementary fibres glued together, bordered in red.**

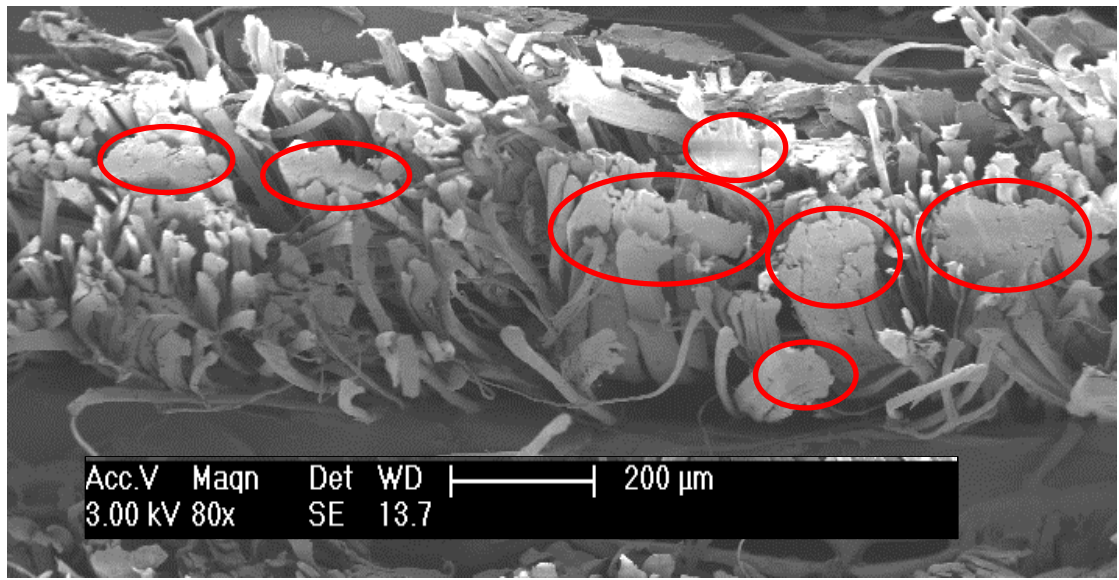

**Figure 11** Highlighted undamaged fabric-2; clusters of elementary fibres glued together, bordered in red

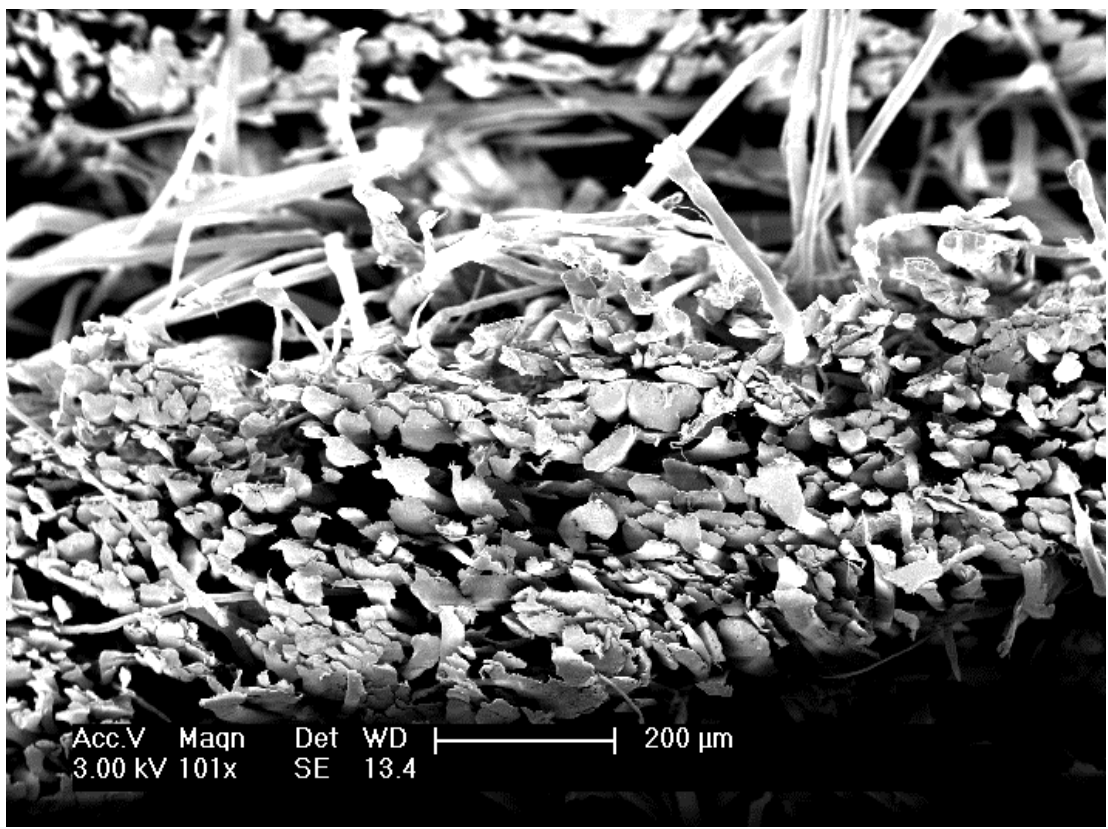

**Figure 12** Damaged fabric specimen-1; the flax fibres are mostly in elementary form.

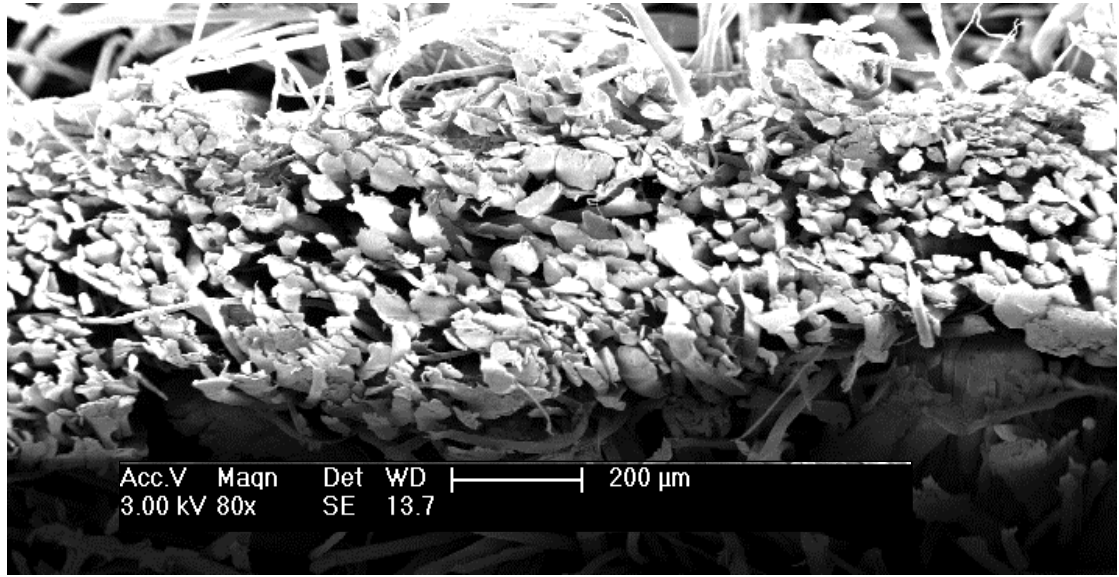

**Figure 13**   Damaged fabric specimen-2; the flax fibres are mostly in elementary form.

Cross-sectional SEM images may help to explain this behaviour. Figure 10 and Figure 11 show the dry flax fabric cross-section before compaction. Figure 12 and Figure 13 show the dry fabric cross-section after compaction.

From studies of flax fibre morphology, it is known that a technical flax fibre is an assembly of multiple elementary fibres. These elementary fibres are embedded within an amorphous matrix of pectin and hemicellulose, overlapping each other to some extent [21, 22]. Upon tensile loading of a technical fibre, the elementary fibres can slide past each other to become individualised [23]. To produce the woven hopsack fabric, warp knitted UD fabric and nonwoven tapes, the initial form of the flax fibre bundle used for this research was hackled sliver. For UD or hopsack production, the slivers go through a drawing operation first, followed by wrap spinning, to convert the fibres into yarn. A fibre histogram was conducted as part of this study to show the length distribution of the fibres from the slivers (the initial hackled slivers) and from the yarns (collected from the UD and hopsack fabrics). The average fibre length of the sliver was 119.4±4.7 mm and from the yarn it was 98.9±4 mm. These lengths are much higher than the average length of the elementary fibres (20 to 50 mm) reported in other research work [21, 24]. The fibre histogram indicated that the fibres remained as bundles of technical fibres in the sliver and in the yarn. The process that converts the slivers into yarns involves a drawing operation which applies axial tension to the axial fibre bundle and individualises them to some extent. Therefore, clusters of elementary fibres become separated from the original technical fibres, i.e. the technical fibres disintegrate into smaller bundles. It can thus be concluded here that the mechanical process can perform the individualisation of the elementary fibres to some extent.

A similar action (disintegration of technical fibre bundles) is likely to occur when the nonwoven tape experiences the compressive load exerted by the steel plates of the Instron machine. Figure 10 and Figure 11 show cross-sectional images of an undamaged fabric cross-section wherein clusters of technical fibres are encircled by red lines. When the fabric is subjected to pressure, the clusters of these fibre bundles may disintegrate into smaller bundles, which increase the quantity of elementary fibres in the fabric. Therefore, under pressure, the thickness of the single ply reduces. Figure 12 and Figure 13 show cross-sectional images of damaged fabric under compression. The presence of the individual fibres is much higher in number in the cross-sections of these damaged fabrics, and this indicates that the technical fibre bundles have been crushed and have disintegrated to some extent under compressive loading. As a result, some additional thickness reduction occurs for a single layer fabric under compressive loading; this has probably happened for both the T170 and the GVT fabrics.

Another particular event can be attributed to the GVT during compaction, that is, the crushing of the glass veils. When a single layer is subjected to compaction between steel plates, the plates crush the surface veils on both sides. But, for a multi-layer stack, the glass veils of the middle layers are not exposed to the steel plates, hence, they remain relatively undamaged, residing as middle layers.

Again, when a multi-layer assembly is stacked with the same fabric, the middle layers experience frictional forces against each other. This frictional force is greater than the frictional force experienced by a single layer fabric against polished steel plates. The cohesive forces between the fibre surfaces of two adjacent layers may act to lock the fibres. Hence, the fibres of one layer cannot settle into the gaps among the fibres from the adjacent layers. The ultimate average thickness per layer calculated from the multi-layer stack may thus be slightly higher than the thickness of an isolated single layer.

This phenomenon is shown in Figure 7. The GVT and T170 fabrics show that the average layer thickness (calculated from the multi-layer stack) is greater than that of the thickness of a single layer. It indicates that nesting did not take place. However, for the hopsack and the UD fabrics, the average layer thickness, is less than the thickness of a single layer and this indicates that nesting has taken place in the multi-layer stack. This signifies that a multi-layer stack of hopsack or UD material possesses the potential to exhibit an increase in fibre volume fraction under compressive pressure during composite formation.

### 3.4 Effect of wettability on macroscopic deformation

During natural fibre composite manufacture, dry fabric is infused with resin which affects the fabric thickness and changes the fibre volume fraction of the preform as reported by various researchers [25-29]. Flax fibre is different from glass and carbon fibre because it has a hydrophilic nature while the mineral fibres exhibit hydrophobic behaviour. In this work, different flax fibre structures were wetted with distilled water to simulate the effect of wettability on thickness and consequently on the fibre volume fraction of various flax fibre structures during transverse compression. A graphical representation is shown in Figure 14. It was observed that in all the structures, thickness reduction was greater in the wet state than for dry fabrics. The reason for this increased thickness reduction can be attributed to the reduced coefficient of friction between the fibre assemblies caused by the lubricating effect of the wetting agent, which make the yarns easier to deform on compaction [30]. Similar behaviour in respect of thickness reduction of lubricated fibres has been reported for different textile fabrics [26, 28, 29].

Another pertinent observation is that, unlike the micro-architecture of glass or carbon fibres, flax fibres contain a lumen, running along the length of the elementary fibres. The role of the lumen inside the flax plant is to assist water transportation for the plants [21, 22, 31]. As a result, when the flax fabric stacks are subjected to resin impregnation, the lumens of the elementary fibres also become impregnated with resin, to some extent [32]. The extent of this effect depends upon the degree of individualisation of the elementary fibres which is, in turn, determined by the fabric manufacturing processes.

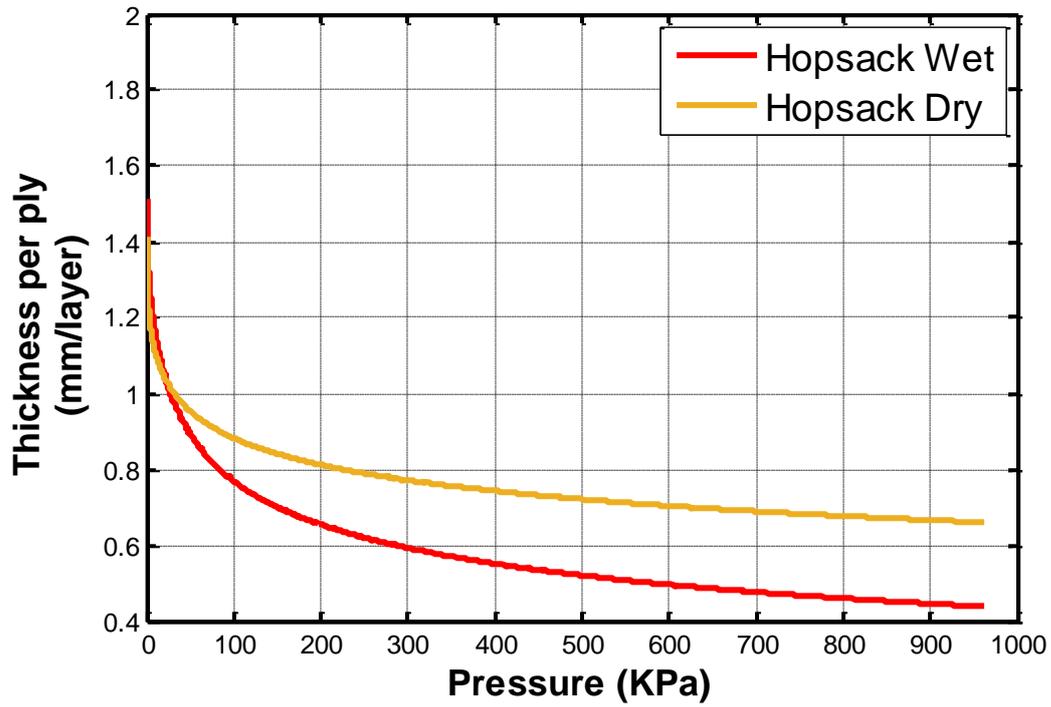

(a)

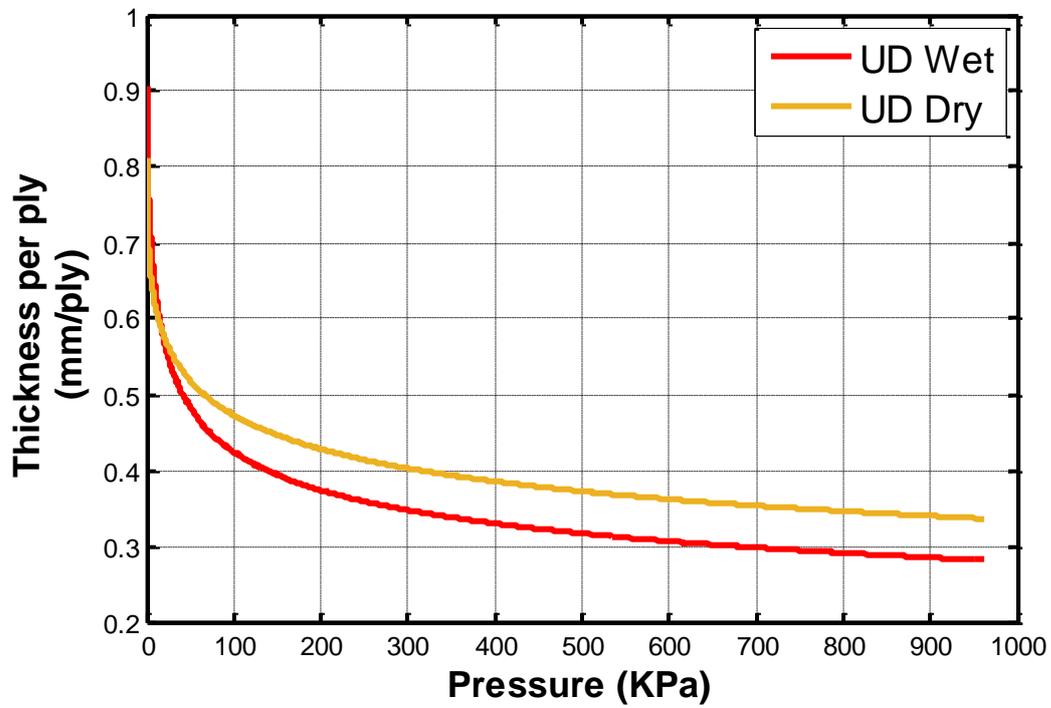

(b)

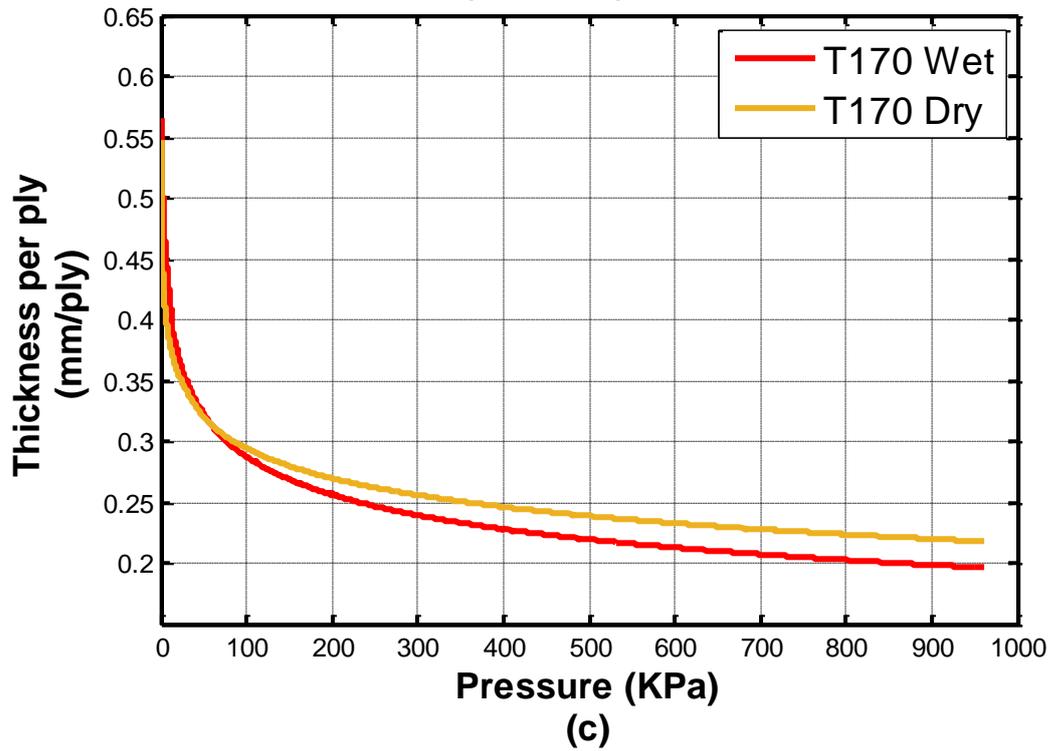

(c)

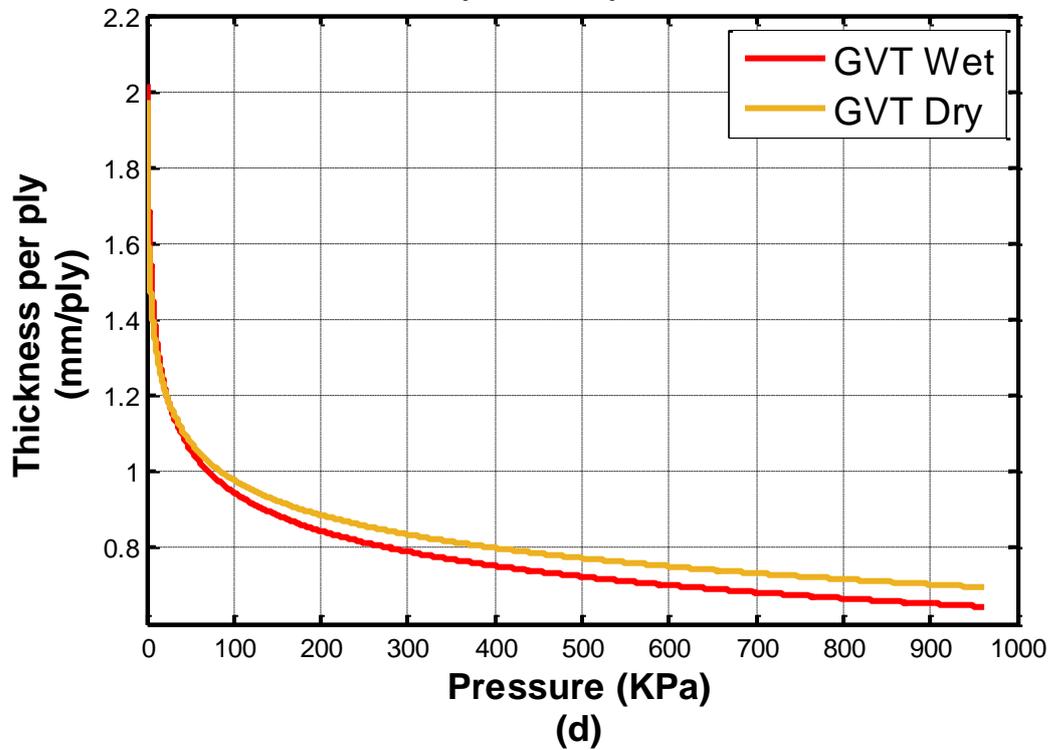

(d)

**Figure 14** Pressure thickness curves of different flax fabric structures for dry and wet fibres.

It is known that the fibre volume fraction of a composite is directly proportional to its constituent fabric plies. Figure 14 shows that the thickness of the wet layers is less than the thickness of the dry layers for all the tested structures, which indicates that appropriate impregnation of resin might be expected to increase the fibre volume fraction of composites.

### 3.5 Effect of ply orientation on macroscopic deformation

During composite manufacturing, plies are place at $0^0$ orientation or $0/90^0$ or at $0/45^0$ depending on the end use of the composites. In this work, different flax fabric architectures have been studied by placing the plies at $0^0$ and $0/90^0$ to study the effect of ply orientation on thickness reduction and on the fibre volume fraction of the preform. The effect of differing ply orientations on GVT, T170 and UD fabrics are graphed in Figure 15.

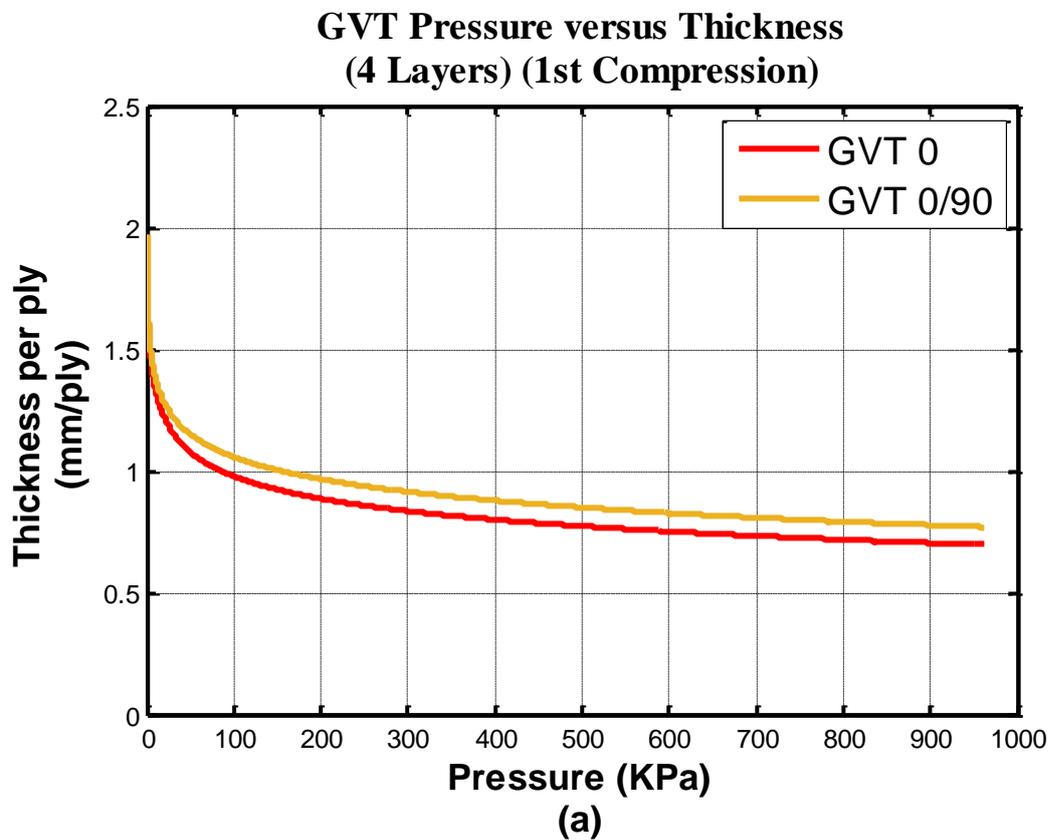

(a)

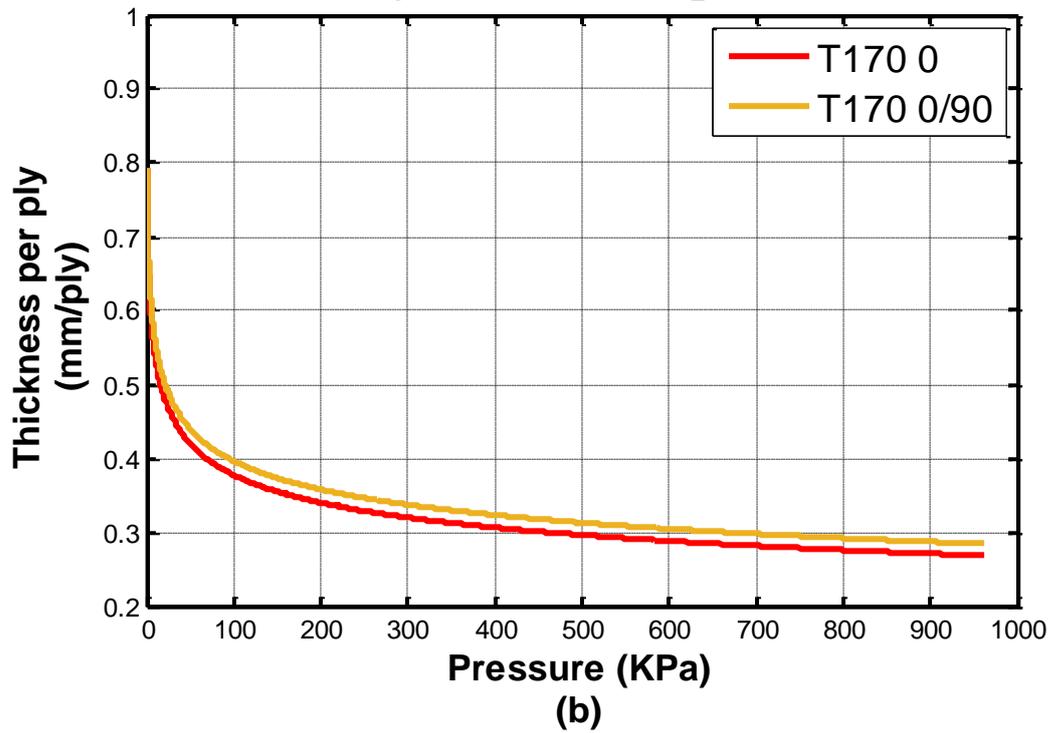

(b)

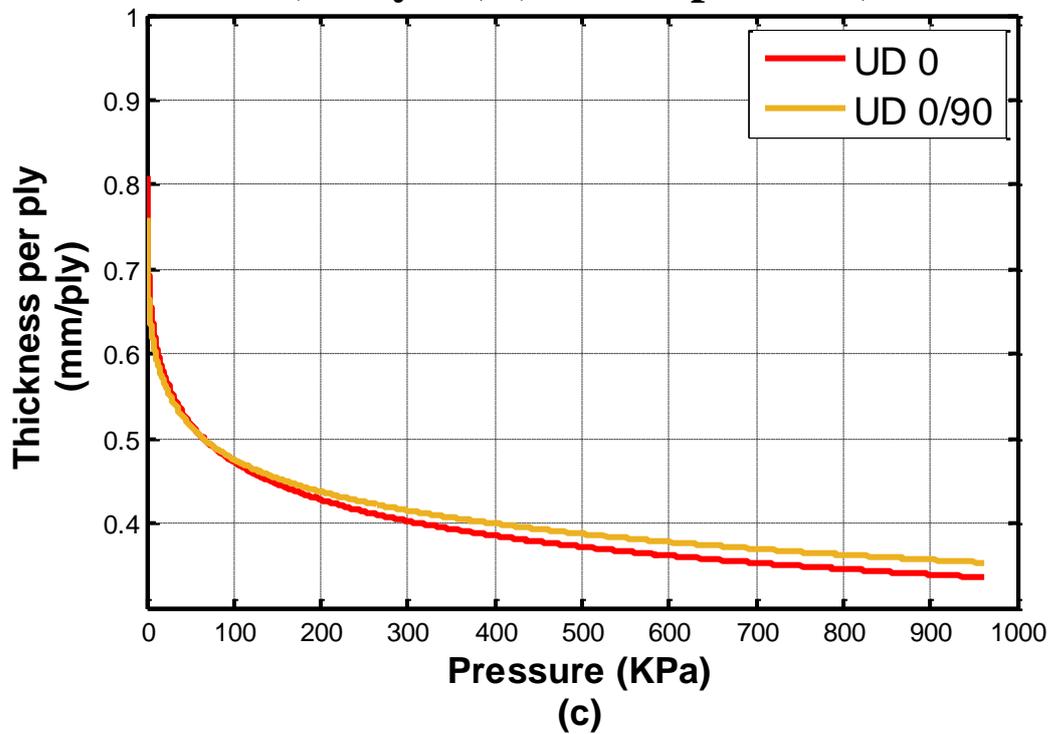

(c)

**Figure 15** Pressure thickness curves of various flax fibre multi-layered structures with $0^0$ and with $0^0/90^0$ ply orientations.

Figure 15 shows all the graphs for the multi-layer stacking of three different structures for cross-ply (0/90°) and unidirectional (0°) ply orientation. Hopsack fabric has been excluded here because the fabric, being a woven structure, is always 0/90°. It may be seen that for all the structures, cross-ply (0°) layup of the fabric plies has yielded lower thicknesses and hence higher fibre volume fractions.

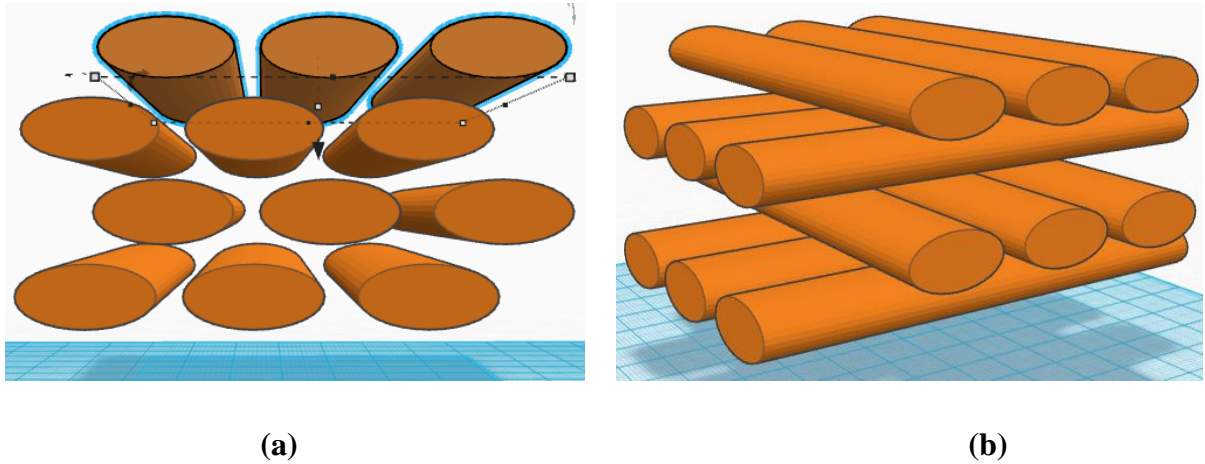

**(a)**            **(b)**

**Figure 16**    **Schematic 3D representation of cross-ply and unidirectional fibre layups.**

When the plies are unidirectionally arranged, the fibres of one layer tend to set into the gaps among the fibres of the adjacent layers, under compression. This settling down of the fibres is easier when the plies are stacked unidirectionally, as shown in Figure 16. But when the layers are stacked in a cross-ply manner, a 90° ply resides between two 0° plies (and vice versa). The crimping of this 90° ply restricts the nesting of the 0° fibres into gaps of the next 0° layer. Also, the 0° fibres can easily nest onto the tow gaps (for fabric) or onto the fibre gaps (for nonwoven tapes; T170 or GVT). In the GVT structure, the restrictions are worsened as a very thin layer of glass veil, attached to both planes of a layer, also hinders the nesting process.

### 3.6 Effect of repeated compression on macroscopic deformation

The need to achieve a high fibre volume fraction for flax fibre composites in different applications makes it necessary to study different compaction methods which can enhance the fibre volume fraction of the composite materials by reducing the preform thickness during forming. It is important to utilise a higher fibre volume fraction for natural fibre composites compared to glass and carbon fibre composites; the reason is that when natural fibre composites are presented as environmentally friendly materials, a high fibre volume fraction maximises the ratio of bio-based material to polymeric material [16]. In this study, the flax fibre structures were compressed under repeated compaction cycles. These multi-layer structures were laid in cross-ply manner, wherein 4 layers of Hopsack (woven preform, hence cross-ply by default) and GVT preforms, 10 layers of T170, and 6 layers of UD were used. Different numbers of layers for the different structures were used to maintain similar fibre volume fractions for each of the composite stacks [shown in the next phase of this current study [32]]. Three compaction cycles were executed on each structure and pressure thickness curves were recorded. It was observed that the fabric thickness decreased during each cycle and the thickness reduction was greatest for the first compaction cycle, whilst the second and third compaction cycles further reduced the thickness of the preform. The degree of further compaction lessened with each successive compression cycle, as seen in Figure 17. Similar behaviour has been reported for glass, carbon [33, 34], and silk fabrics [16]; in all these publications fabric thickness reduction was observed with cyclic compression and the initial compaction resulted in the greatest thickness reduction. Hence, it can be concluded that to obtain a higher fibre volume fraction, flax fabric plies should be compressed at least once. It has also been shown, in section 3.3, that relatively high compressive pressure applied by two steel plates upon a single layer splits the technical fibre bundles into smaller bundles, i.e. more individualisation takes place. Stuart et al. (2006) have reported that a treatment of flax fibres with ethylene diamine tetraacetic acid (EDTAA) breaks down the pectin bonds of the technical fibres that hold together the elementary fibres as a bundle. As a result, during the creation of composite structures, more fibre surface becomes exposed for matrix bonding in comparison with the original technical fibre bundle (because of the lower aspect ratio of the technical fibres). Ultimately, the tensile strength of flax/epoxy composites increases by 50%, although no significant change of modulus has been reported [35]. Nevertheless, further study is required to quantify the extent of the individualisation of the elementary fibres caused by compressive loading applied by steel plates.

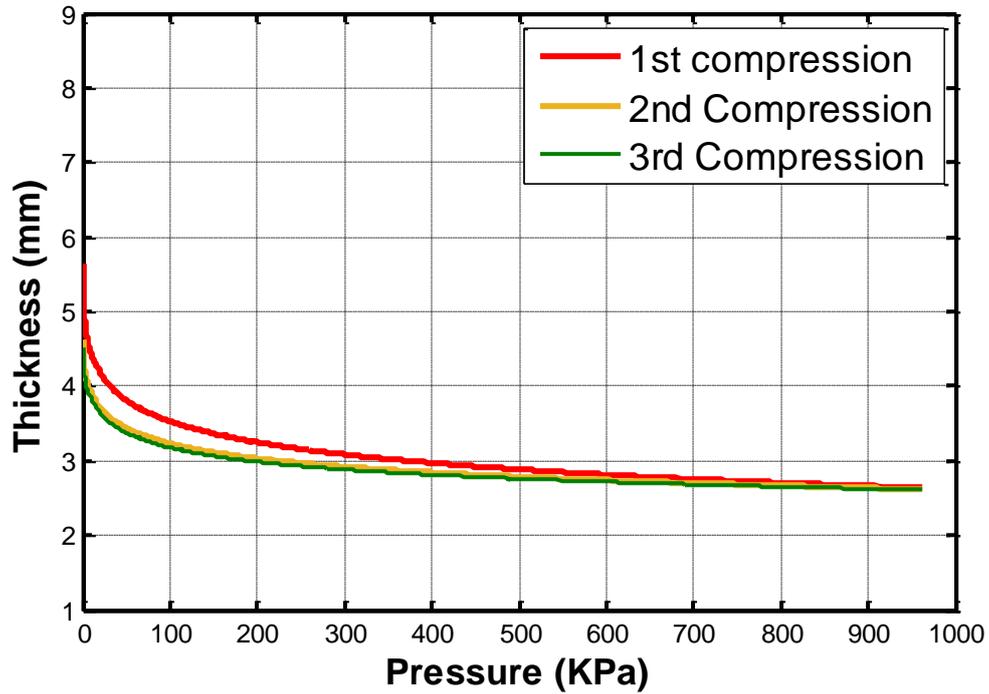

(a)

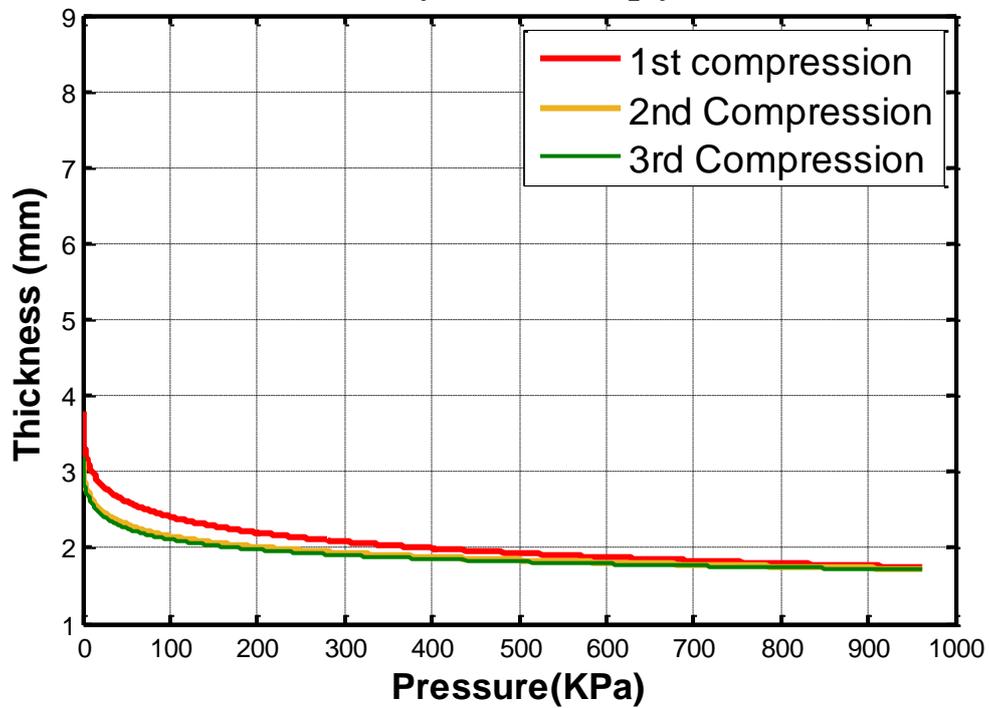

(b)

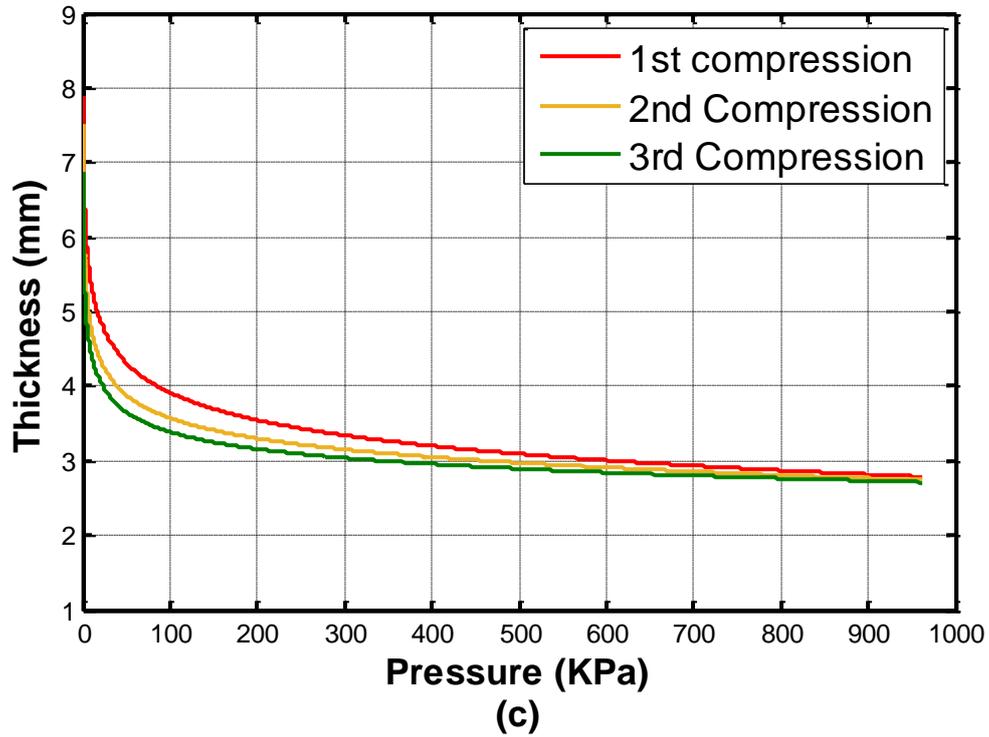

(c)

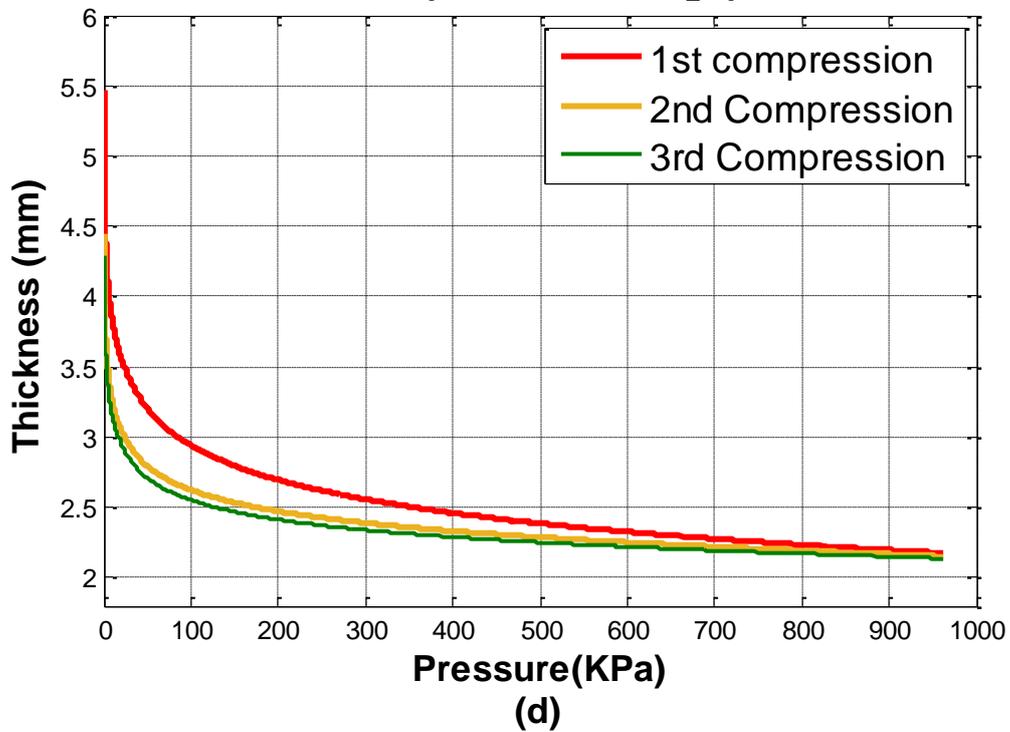

(d)

**Figure 17** Pressure thickness curves for various flax fibre structures subjected to repeated compaction.

## 3.7 Relevance of nesting factor with fibre volume fraction

The nesting factor has been calculated for the all the four structures examined during the course of this research and a description has been given in section 3.2. A lower nesting factor indicates more concentrated nesting, which in turn implies that the fibre volume fraction of composites made from that stack of plies will be greater. For a stack of fabric plies, more intense nesting of the reinforcement ensures that the overall stack thickness is reduced, and hence the volume of the composite diminishes without changing the volume of the constituent fibres. So, a higher fibre volume fraction can be achieved with higher nesting. Figure 18 illustrates this concept using a schematic diagram. Figure 19 summarises the nesting factors for various structures at different levels of applied pressure.

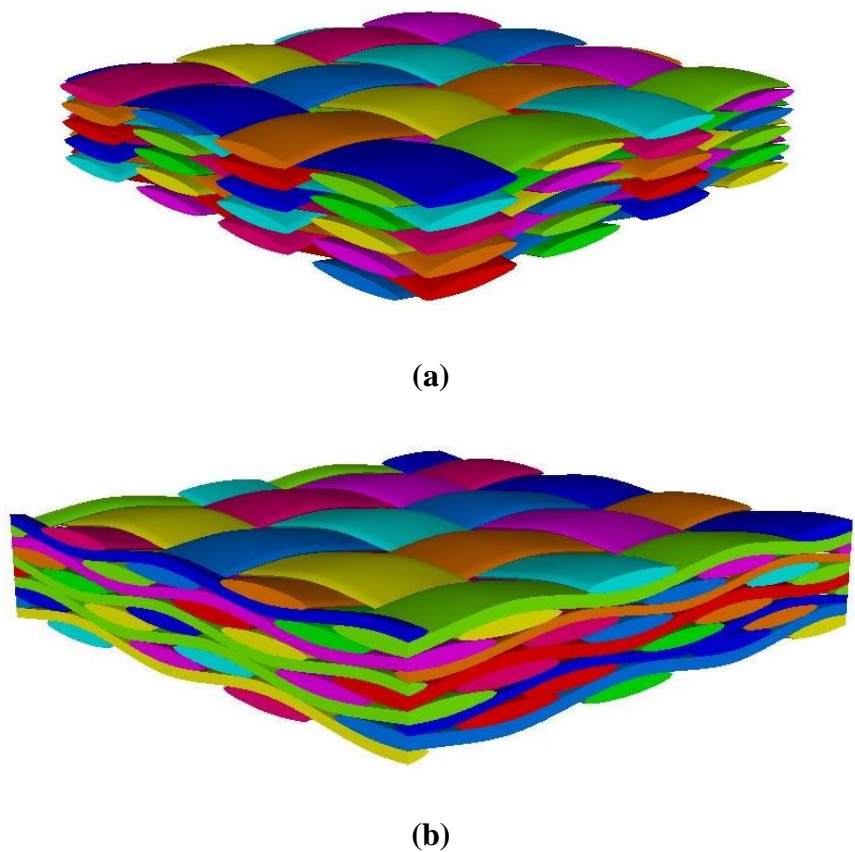

**Figure 18   Schematic diagram showing a fabric stack, (a) without shifting and nesting; and (b) after shifting and nesting of fabric layers.**

It can be seen from Figure 18 (a) that the warp yarns are laid upon each other, whereas in Figure 18 (b), the yarns have settled into the gaps between yarns of the adjacent layers. Therefore, the overall volume of the gaps between the yarns is greater in Figure 18 (a) than in Figure 18 (b). Thus, it is expected that a higher fibre volume fraction will result from the increased level of nesting. This improved nesting (and smaller nesting factor) lead to a reduction in the number and the extent of resin rich areas, which is generally advantageous for the engineering properties of fibre-reinforced composites. From Table 4, it may be concluded that multi-layer stack thickness will decrease with increasing externally applied pressure (Figure 6). Likewise, the nesting of layers is improved by application of pressure (Figure 19) and the fibre volume fraction also increases (Figure 20).

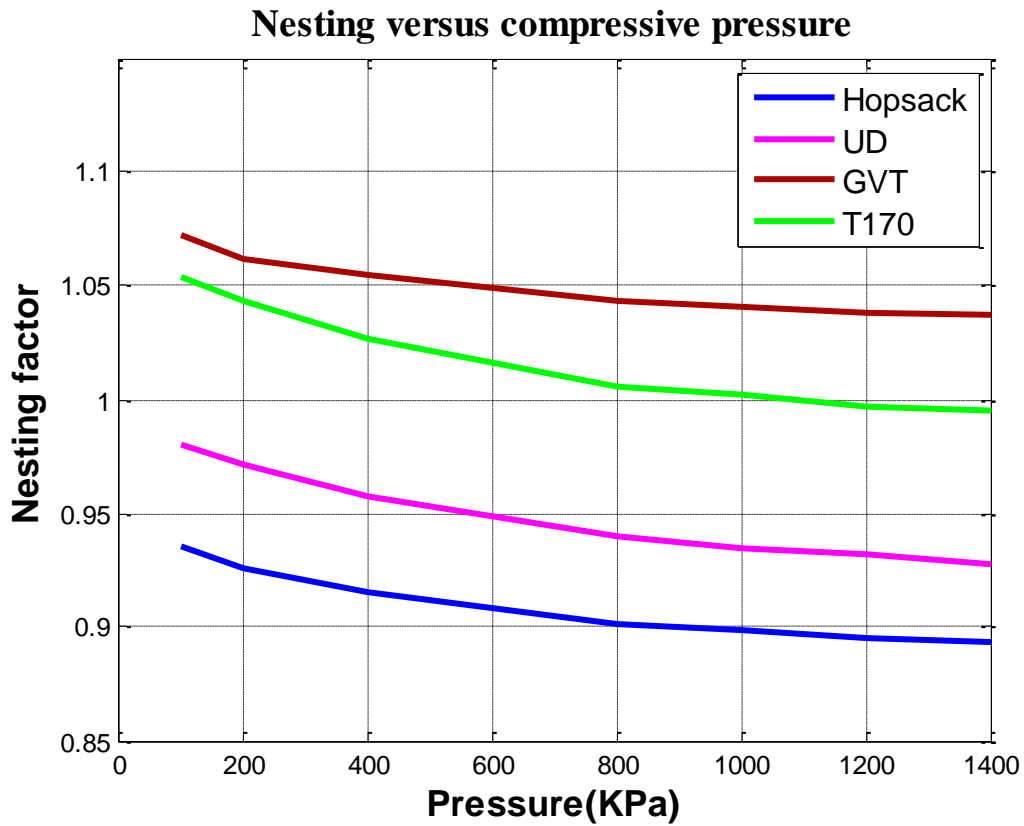

**Figure 19   Variation of nesting factor as a function of pressure.**

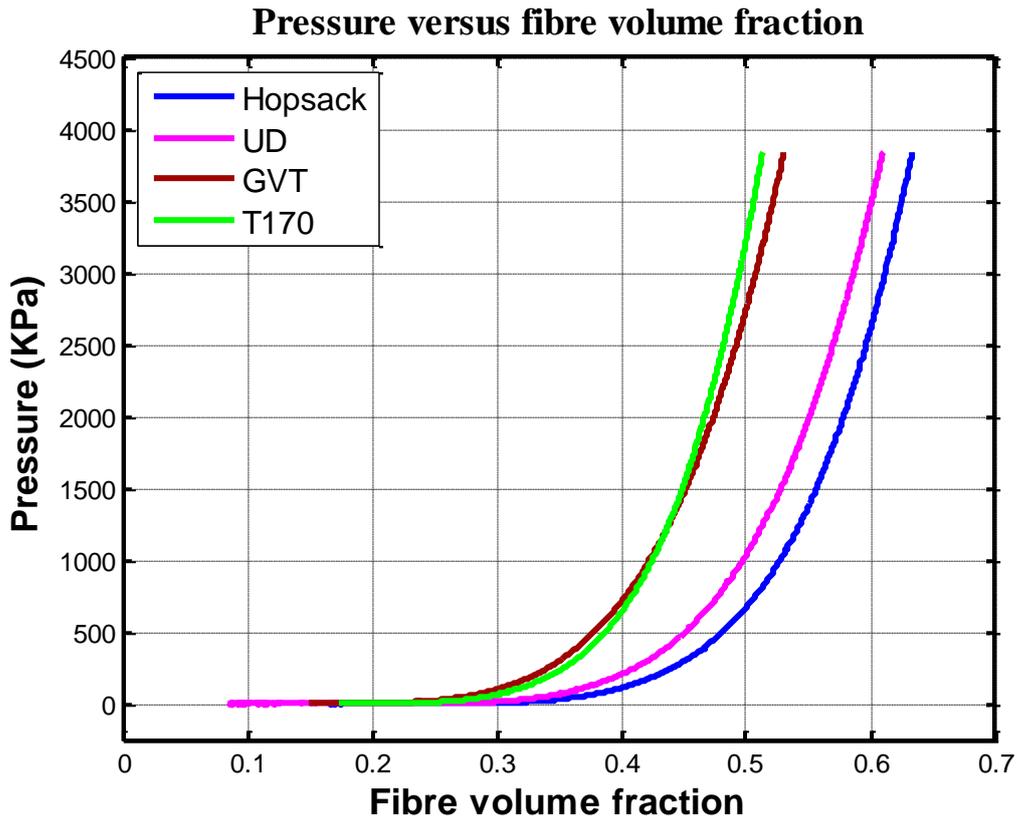

**Figure 20** Pressure versus fibre volume fraction by mathematical method.

### 3.8 Fibre volume fraction analysis

Three methods namely: theoretical calculations for composite laminate, dry preform and ImageJ analysis have been employed in this study to determine fibre volume fraction. The results from these three methods are summarised in Figure 24, which indicates that they are in very good agreement with each other. Among these methods, image analysis can sometimes be less reliable because impurities on the cross-sectional surfaces of the specimens can result in misinterpretation by the analytical software. Unlike the homogeneous cross-sectional images generated by glass fibre composites as shown in Figure 21 (a), flax reinforced composites, as seen in Figure 21 (b), contain lumens, running through the elementary fibres. It becomes difficult to precisely distinguish the fibre, resin, and hollow areas using computerised image analysis as the thresholding techniques are defeated by the complex structure of the flax fibres. The magnitude of the problem is exacerbated when the image is captured at lower magnifications, below 200. To avoid such issues, Figure 22 (b) has been captured at 400 times magnification where all the elements such as fibre, matrix, and lumen void can be identified distinctly. However, when an image such as Figure 23 (a), is captured at 100 times magnification, the ImageJ software cannot distinguish the different elements as precisely as for a highly magnified image.

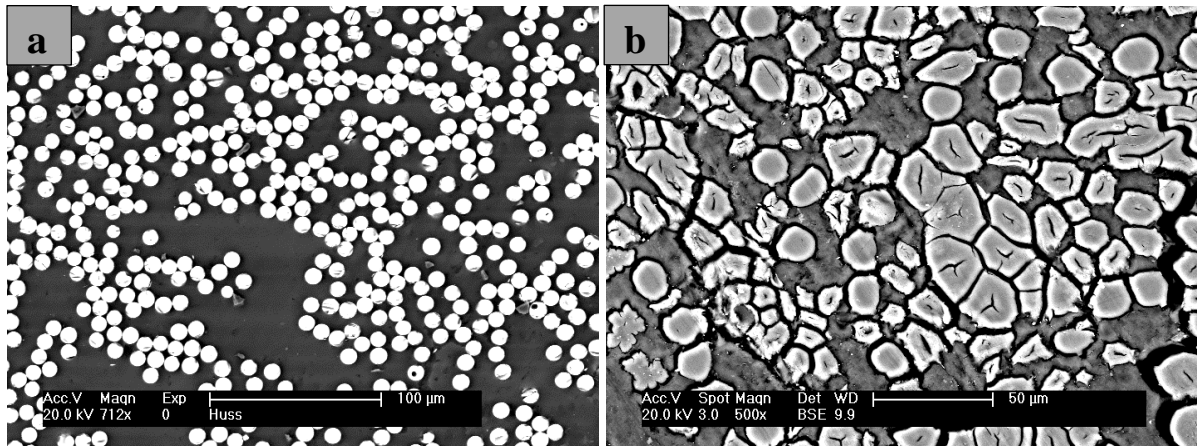

**Figure 21  Cross-sectional images of composites which contrast a glass reinforced composite structure, (a); witha flax reinforced composite, (b).**

### 3.8.1 Fibre volume fraction by image analysis

Cross-sectional images of flax-based composites were initially captured using scanning electron microscopy (SEM). Each image was then analysed by using an analytical tool known as 'ImageJ'. Using ImageJ, each cross-sectional image was analysed by applying a binary thresholding tool. In thresholding, the greyscale images obtained from SEM were reduced to binary images by introducing an optimal threshold. During thresholding, pixels from the image that represented a specific object were extracted. Thresholding generates a binary image in which the foreground regions are indicated by one single intensity and the background regions appears as different, contrasting intensities. After binary thresholding of each image, the fibres in the foreground were converted into black and the other elements of the picture, such as resin and lumen were converted into white, as demonstrated in Figure 22. Figure 22 (a) shows a greyscale SEM image of the cross-section of a T170 composite sample. Figure 22 (b) shows the stack contrast image after thresholding. This binary thresholding has been performed in such a way that only the light grey flax fibres have been converted into a black colour. The hollow spaces comprising the lumen appear in white. These hollow spaces have been encircled on both the images, showing that the thresholding has been able to differentiate the fibres fully from the resin, and the hollow spaces. The images after thresholding were 'outlined' to show just the flax fibres in the image, as seen in Figure 22 (c). After thresholding, the image was then analysed by a 'Particle Analysis' tool where the area of the black portion was calculated as a ratio of the total area. Figure 22 contains the scale bar; this is not retained in the images used to analyse the fibre volume fractions of a composite because the elements of scale bars are also counted by the 'Particle Analysis' tool. The analysis of Figure 22 (b) finds that a 46.08% area is occupied by fibres and an example of the process of using ImageJ is illustrated in Figure 23.

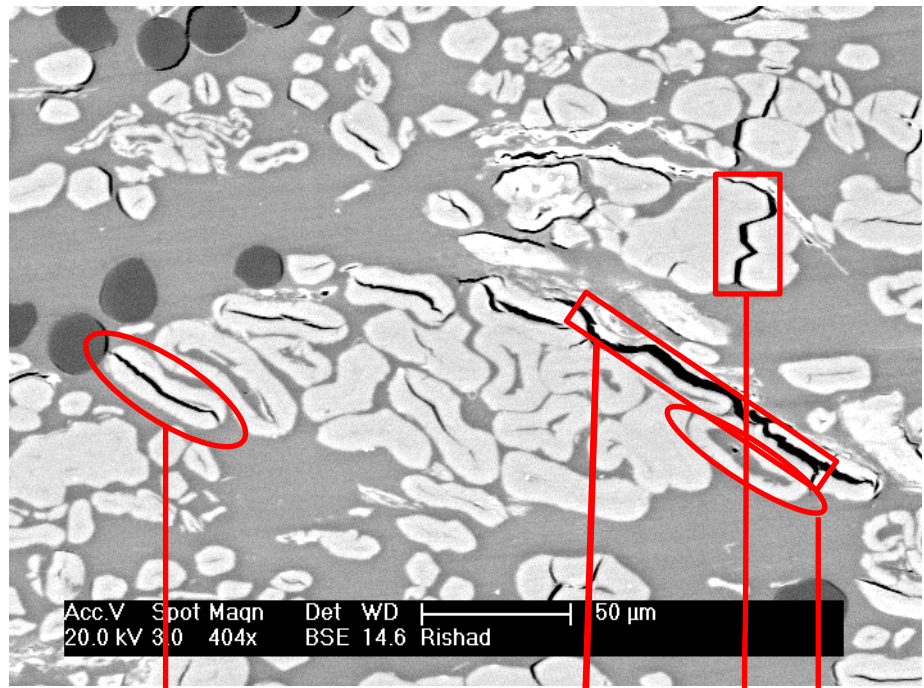

(a)

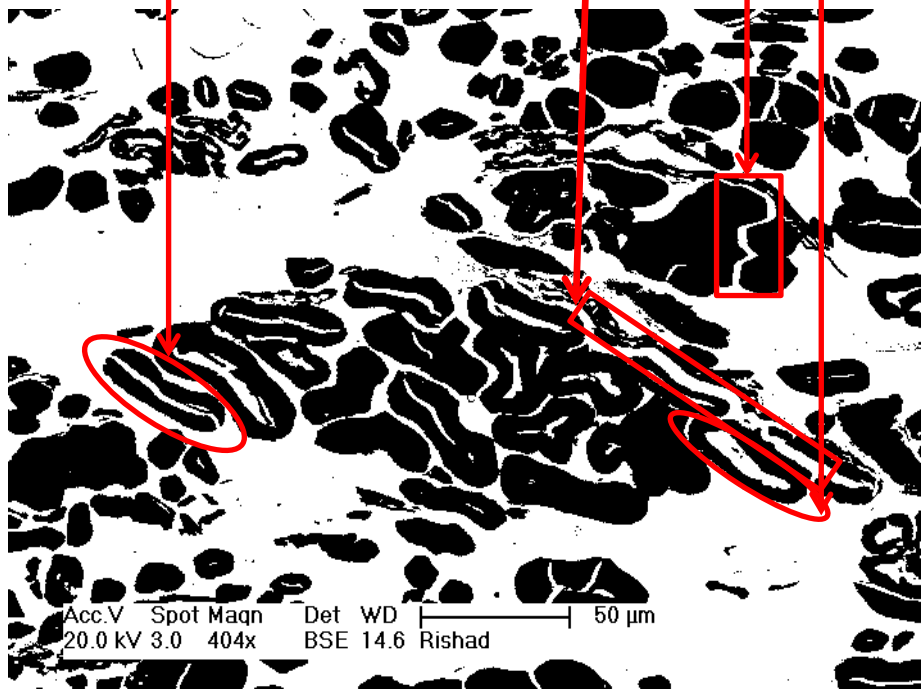

(b)

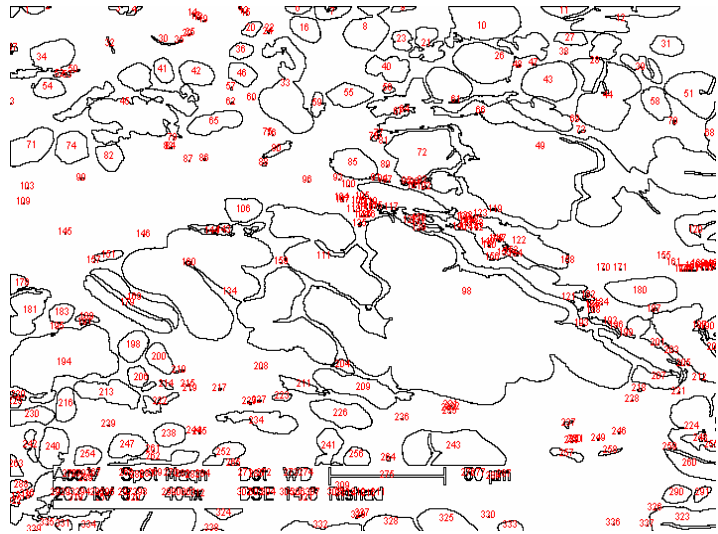

**(c)**

**Figure 22  Image of a T170 composite cross-sectional surface; (a) greyscale image of fibres before thresholding; (b) high contrast image after thresholding; and (c) Outlined image of fibres.**

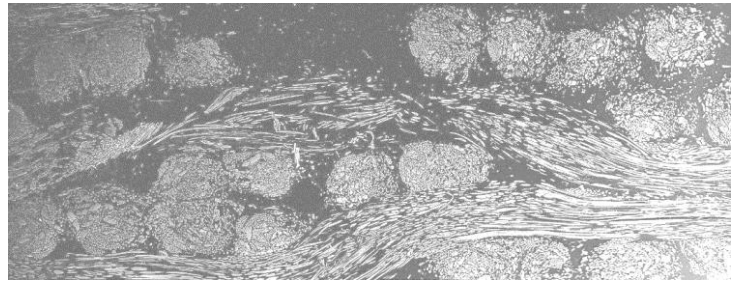

**(a)  SEM cross section of Hopsack composites: step 1 of image analysis.**

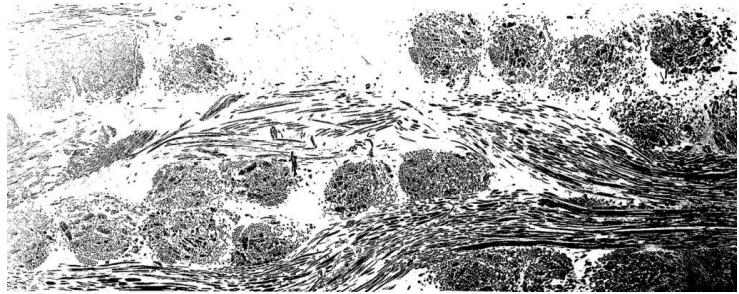

**(b)  Binary thresholding using imageJ: step 2 of image analysis.**

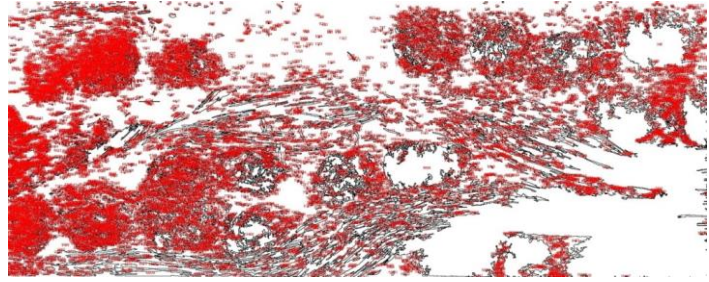

**(c) Particle analysis and area measurement using ImageJ: step 3 of image analysis.**

**Figure 23  ImageJ used to calculate fibre volume fraction for hopsack composite.**

The fibre volume fraction calculated using ImageJ was 0.38; 38%.

**3.8.2 Theoretical measurement of fibre volume fraction of composite laminate**

To estimate fibre volume fraction using a theoretical approach, firstly, the areal density of each different fabric structure was measured. Then the weight of each composite test coupon (cut according to standard) was measured using a precision weighing machine. Subsequently, all the three dimensions of the test coupon were measured by using a digital micrometer. The volume of fibres in different structures was calculated by the following expression:

$$Volume\ of\ fibre, V_f = \frac{mass\ of\ flax\ \times \%\ of\ flax\ in\ the\ fabric}{density\ of\ fibre} + \frac{mass\ of\ polyester\ \times \%\ of\ polyester\ in\ the\ fabric}{density\ of\ polyester}$$

Table 1 summarises the specification. Finally the fibre volume fraction of composite specimens was calculated by diving fibre volume to total volume of composite laminate.

**3.8.3 Theoretical measurement of fibre volume fraction of dry preform**

The fibre volume fraction of dry preform samples was calculated by the following expression:

$$v_f = \frac{n \times k}{t_c \times \rho_f}$$

where:

$n = number\ of\ fabric\ layers$

$k = areal\ density\ in\ \frac{g}{cm^2}$

$$t_c = thickness\ of\ fabric, in\ cm$$

$$\rho_f = density\ of\ fabric, in\ \frac{g}{cm^3}$$

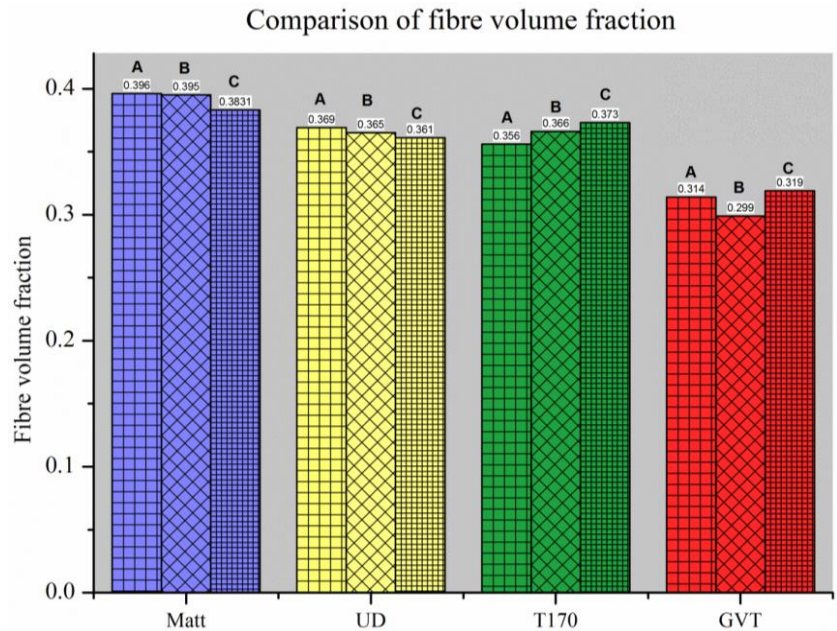

**Figure 24** Comparison of fibre volume fractions, measured by different techniques, theoretical calculati for composite laminate (A), dry preform (B) and by image analysis (C)

## 4. Conclusions

A number of flax fibre structures have been investigated for compaction behaviour during composite forming, in order to study the thickness changes of these fabrics up to a pressure of 10 bars, covering a range of composite manufacturing processes including vacuum infusion, autoclave and resin transfer moulding. The fabrics were studied in single and multi-layer, dry and wet state, under different loading cycles and when arranging the fabric plies in different orientations. Nesting of the layers was calculated for single plies and for multi-layer stacks of dry fabrics. It was observed that the thickness of all the flax fabrics reduced in both single layers and multi-layer stacks, and followed typical pressure-thickness curves. The thickness reduction of hopsack and unidirectional fabrics was greater in multi-layer stacks, which was attributed to nesting of the layers. The thickness reduction of T170 (nonwoven tape) and GVT (veiled) structures was greater in the single layers compared to the multi-layer stacks, which was attributed to the higher deformation of the single layers between the low friction polished steel compression plates and the absence of nesting in the multi-layer stacks. As a result, nesting factors were greater than 1 for the nonwoven tapes, which can be attributed to the higher frictional forces amongst the fibres in adjacent layers in a multi-layer stack. Other than frictional force, crushing of technical fibre bundles into smaller clusters of fibres has been observed as a reason for the generation of nesting factors higher than 1. More pronounced thickness reduction was also observed for wet fabrics compared to dry fabric compaction, which was due to reduction in the frictional forces between the fibres, as water acts as a source of lubrication on the fibre surfaces. Ply orientation also affected the thickness reduction. Repeated compaction of the flax fabrics resulted in greater degrees of thickness reduction though the effect of repetitive compaction decreased as the number of cycles increased. The smallest thickness reduction in the $3^{rd}$ cycle was due to the permanent deformation of the fibres. Further study is required to quantify the crushing of technical fibres by compressive loading, which is potentially beneficial for the composites' mechanical properties.